\documentclass[useAMS,usenatbib]{mn2e}
\usepackage{graphicx,amsmath,epsfig,rotating,fleqn} 

\newcommand{\figdir}
  {./}
\newlength{\narrowfigurewidth}
\newlength{\figurewidth}
\newlength{\widefigurewidth}
\newcommand{\unit}[1]
  {{\mbox{\rm\,\,#1}}}
\newcommand{\percent}
  {per~cent}

\newcommand{\etc}
  {etc.}
\newcommand{\etal}
  {et al.}
\newcommand{\eg}
  {e.g.}
\newcommand{\cf}
  {cf.}
\newcommand{\ie}
  {i.e.}
\newcommand{\diff} 
  {{\rmn{d}}}
\newcommand{\chisq}
  {\chi^2}
 \newcommand{\vidrih}
  {V07}
\newcommand{\vect}[1]
  {\mbox{\boldmath ${#1}$}}

\newcommand{\eq}[1]
  {Eq.~(\ref{equation:#1})}
\newcommand{\eqs}[1]
  {Eqs~(\ref{equation:#1})}
\newcommand{\sect}[1]
  {Section~\ref{section:#1}}
\newcommand{\sects}[1]
  {Sections~\ref{section:#1}}
\newcommand{\tabl}[1]
  {{\mbox Table~\ref{table:#1}}}
\newcommand{\tabls}[1]
  {{\mbox Tables~\ref{table:#1}}}
\newcommand{\fig}[1]
  {Fig.~\ref{figure:#1}}
\newcommand{\figs}[1]
  {Figs.~\ref{figure:#1}}

\newcommand{\prob}
  {P}
\newcommand{\prior}
  {\upi}
\newcommand{\posterior}
  {P}
\newcommand{\evidence}
  {E}

\newcommand{\likelihood}
  {L}
  
\newcommand{\kms}
  {\unit{km} \unit{s}^{-1}}

\newcommand{\dtdt}
  {\dot{T}_{\rmn{eff}}}
\newcommand{\thin}
  {{\rmn{thin}}}
\newcommand{\thick}
  {{\rmn{thick}}}
\newcommand{\halo}
  {{\rmn{halo}}}
\newcommand{\core}
  {{\rmn{core}}}
\newcommand{\nband}
  {N_{\rmn{b}}}
\newcommand{\band}
  {b}
\newcommand{\nparam}
  {N_{\rmn{p}}}

\newcommand{\parameters}
  {\vect{\theta}}
\newcommand{\parametersobs}
  {\hat{\vect{\theta}}}
\newcommand{\parameter}
  {\theta}
\newcommand{\data}
  {\vect{d}}

\newcommand{\dist}
  {D}

\newcommand{\vel}
  {\vect{V}}
\newcommand{\distobs}
  {\hat{D}}
\newcommand{\Sun}
  {\odot}
\newcommand{\msun}
  {M_{\Sun}}
\newcommand{\hydrogen}
  {{\rmn{H}}}
\newcommand{\helium}
  {{\rmn{He}}}

\newcommand{\teff}
  {T_{\rmn{eff}}}
\newcommand{\teffobs}
  {\hat{T}_{\rmn{eff}}}
\newcommand{\teffobsgrid}
  {\hat{T}_{\rmn{eff,grid}}}
\newcommand{\teffminrj}
  {6 \times 10^4\unit{K}}
\newcommand{\logg}
  {\log{\mbox{\em{g}}}}
\newcommand{\loggpar}
  {g}

\newcommand{\mappobs}
  {\hat{m}}
\newcommand{\mabs}
  {M}

\newcommand{\bestfit}
  {best-fit}
\newcommand{\Bestfit}
  {Best-fit}
\newcommand{\nhalo}
  {34}
\newcommand{\neisen}
  {17}
\newcommand{\nultracool}
  {24}
\newcommand{\vtan}
  {V_{\rmn{tan}}}
\newcommand{\vtanobs}
  {{\hat{V}_{\rmn{tan}}}}
\newcommand{\vtanobsgrid}
  {{\hat{V}_{\rmn{tan, grid}}}}
\newcommand{\loggobs}
  {{\log \hat{g}}}

\newcommand{\uprime}
  {\mbox{$u$}}
\newcommand{\gprime}
  {\mbox{$g$}}
\newcommand{\rprime}
  {\mbox{$r$}}
\newcommand{\iprime}
  {\mbox{$i$}}
\newcommand{\zprime}
  {\mbox{$z$}}
\newcommand{\umg}
  {\mbox{\uprime$-$$\gprime$}}
\newcommand{\gmr}
  {\mbox{\gprime$-$$\rprime$}}

\title
  [Photometric constraints on white dwarfs]
  {Photometric constraints on white dwarfs 
     and the identification of extreme objects}

\author
  [D.\ J. Mortlock \etal]
  {Daniel J.\ Mortlock$^{1,2}$\thanks{E-mail: mortlock@ic.ac.uk},
  Hiranya V.\ Peiris$^{2,3}$
  and
  \v{Z}eljko Ivezi\'{c}$^4$
\vspace{7mm}\\
$^1$Astrophysics Group, Imperial College London, Blackett Laboratory,
  Prince Consort Road, London, SW7 2AZ, U.K.\\
$^2$Institute of Astronomy, Madingley Road, Cambridge CB3 0HA, U.K.\\
$^3$Kavli Institute for Cosmological Physics
  and Enrico Fermi Institute, University of Chicago, Chicago, IL 60137, USA\\
$^4$Department of Astronomy, University of Washington,
  Seattle, WA 98195, USA}

\begin{document}

\date{Received 2008 October 27}

\pagerange{\pageref{firstpage}--\pageref{lastpage}} \pubyear{2008}

\maketitle

\label{firstpage}

\begin{abstract}
It is possible to reliably identify
white dwarfs (WDs) without recourse to spectra,
instead using photometric and astrometric measurements
to distinguish them from Main Sequence stars and quasars.
WDs' colours can also be used to infer their intrinsic properties
(effective temperature, surface gravity, \etc),
but the results obtained must be interpreted with care.
The difficulties stem from the existence of a solid angle degeneracy,
as revealed by a full exploration of the likelihood, 
although this can be masked if a simple \bestfit\ approach is used.
Conversely, this degeneracy can be broken if a Bayesian approach is adopted,
as it is then possible to utilise the prior information
on the surface gravities of WDs implied by spectroscopic fitting.
The benefits of such an approach are particularly strong
when applied to outliers, such as the candidate
halo and ultra-cool WDs identified by Vidrih \etal\ (2007).
A reanalysis of these samples confirms their results for the 
latter sample 
but suggests that that most of the halo candidates are thick disk WDs
in the tails of the photometric noise distribution.
\end{abstract}



\section{Introduction}
\label{section:intro}

White dwarfs (WDs) are sufficiently long-lived that they can be 
used to measure the Galaxy's 
age (\eg, \citealt{Fontaine:2001, Fontaine:2005} and references therein)
and
star-formation history (\eg, \citealt{Wood:1992}),
and can even be used as cosmological probes 
(\eg, \citealt{Bergeron:1997, Leggett:1998, Harris_etal:1999, Hodgkin:2000,
  Gates_etal:2004, Harris_etal:2008}).
WDs in the Galaxy's halo
have also been invoked as potential dark matter candidates
(\eg, \citealt{Oppenheimer:2001a,Oppenheimer:2001b}),
although these claims are the subject of considerable debate
(\eg, \citealt{Reid:2005} and references therein).


All such investigations rely on being
able to constrain a WD's properties:
its distance, $\dist$,
and its intrinsic parameters
(effective temperature, $\teff$; 
surface gravity,
$\loggpar$\footnote{The convention of characterising surface gravity, 
$\loggpar$, by 
$\logg = \log(g_{\rmn{cgs}}) = \log[g / (1 \unit{cm} \unit{s}^{-2})]$ 
is followed here; the relationship to the SI equivalent is 
$\log(g_{\rmn{SI}}) = \log(g_{\rmn{cgs}}) - 2$.}; 
absolute magnitude, $M_{\rmn{bol}}$; radius, $R$;
atmospheric composition; \etc).
The distances to a number of the closest WDs (with $\dist \la 100\unit{pc}$)
have been determined unambiguously by parallax measurements 
(\eg, \citealt{Bergeron:2001}), 
and the {\em{Gaia}} satellite\footnote{See the {\em{Gaia}} web-site at 
{\tt{http://www.esa.int/science/gaia}}.} \citep{Perryman:2001} will 
eventually measure parallaxes for all Galactic sources brighter than 
$r \simeq 20$ (\eg, \citealt{Torres:2005,Jordan:2007}).
In most cases, however, 
the characterisation of WDs is 
necessarily 
more model-dependent, with spectroscopic or photometric data used 
to infer the physical properties of WDs.
WD spectra encode sufficient information to infer 
$\teff$ and $\logg$ to a relative accuracy of $\la 10$ per cent (\eg, 
\citealt{Bergeron:1992,McCook:1999,Harris_etal:2003,Kleinman:2004,Eisenstein_etal:2006}),
and the agreement between the various semi-independent models
(\eg, \citealt{Bergeron:1995,Finley:1997,Fontaine:2001,
  Bergeron:2001,Bergeron:2002,Holberg:2006})
gives grounds for confidence in the resultant parameter estimates.
Perhaps the most impressive demonstration of 
spectroscopic fitting 
has been the largely automated analysis of $\sim 10^4$ 
Sloan Digital Sky Survey (SDSS; \citealt{York:2000}) WDs
by \cite{Eisenstein_etal:2006}.
This confirmed that most WDs have $\logg \simeq 8$
(\cf\ \citealt{Bergeron:1992,Liebert:2005}),
but also that the inferred surface gravities and masses
of cool ($\teff \la 10^4\unit{K}$) WDs can be biased high  
(\cf\ \citealt{Bergeron:1990,Kleinman:2004,Eisenstein_etal:2006}).
Interestingly, such a bias is not seen in purely photometric fits 
\citep{Engelbrecht:2007}.
Combined with the fact that it is possible to 
reliably identify WDs using only photometric and astrometric data 
(\eg, \citealt{Harris_etal:2006,Kilic:2006}),
it is of great importance to understand what constraints can be
placed on a WD's properties from photometric data alone.

\begin{figure*}
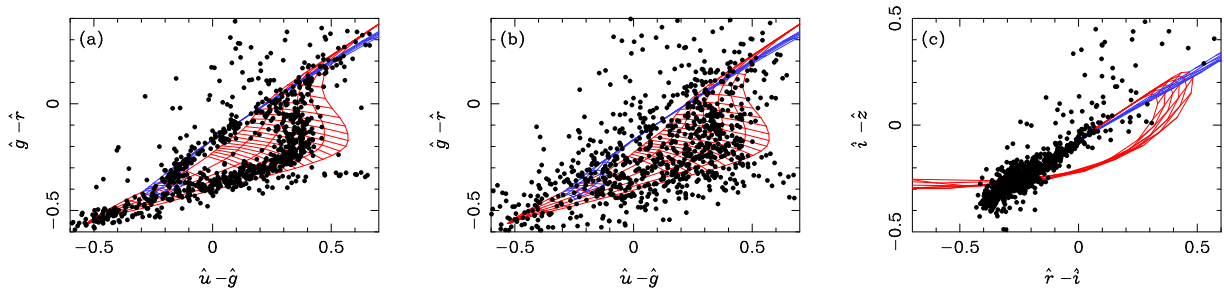

\includegraphics[width=\narrowfigurewidth]{\figdir ugr.eps}
\includegraphics[width=\narrowfigurewidth]{\figdir ugr_noisy.eps}
\includegraphics[width=\narrowfigurewidth]{\figdir riz.eps}
\caption{Two colour diagrams for a sample of non-variable
point--sources in SDSS Stripe~82 from Ivezi\'{c} \etal\ (2007).
In (a) and (c) the sources have relative photometric errors of 
$\sim 0.01$ in all bands;
in (b) the errors have been inflated to be $\sim 0.05$.
The Fontaine--Bergeron model grids for H (red) and He (blue) WDs are also shown,
the temperature going from $\teff = 110000 \unit{K}$ at the bottom-left
down to $\teff \simeq 6000 \unit{K}$ at the top-right,
with the exception of the coolest H models in (c).}
\label{figure:twocolour}
\end{figure*}

The most optimistic case is illustrated in \fig{twocolour}~(a),
in which colour measurements of non-variable point--sources 
generated from multi-epoch SDSS Stripe~82 data by \cite{Ivezic:2007}
are compared to predicted photometry generated by 
G.\ Fontaine and P.\ Bergeron\footnote{Machine-readable versions of the
Fontaine--Bergeron H and He atmosphere WD model grids,
giving broad-band magnitudes for a range of temperatures
and surface gravities,
are currently available from
{\tt{http://www.astro.umontreal.ca/$\sim$bergeron/CoolingModels/}}.
These are used for all the results presented here
as the focus is on random statistical uncertainties;
hence the the systematic variations between 
Bergeron's models and the semi-independent models of, \eg,
\cite{Bergeron:1995},
\cite{Finley:1997},
\cite{Fontaine:2001}
and \cite{Holberg:2006} are not explored.}.
The sources shown have relative photometric errors of
$\sim 0.01$ in all bands,
and are mainly WDs,
although there is some contamination at high
$\gmr$ from quiescent quasars \citep{Sesar:2007}
and unresolved pairs of white and red dwarfs \citep{Smolcic:2004}.
Comparison of the observed WD loci with the models in \fig{twocolour}~(a)
reveals not only an obvious separation between the H and He WDs,
but also clear differentiation in $\teff$.
The surface gravity is well constrained 
for H atmosphere WDs with 
$10^4 \unit{K} \la \teff \la 2 \times 10^4 \unit{K}$
(\ie, $\umg \simeq 0.3$ and $\gmr \simeq -0.2$).
Unfortunately, this level of characterisation,
particularly in $\logg$,
relies on Stripe~82's 
near-ideal combination of photometric precision and wavelength coverage.
\fig{twocolour}~(b) shows the results of 
increasing the relative photometric errors of the Stripe~82 data to $\sim 0.05$:
the H and He loci are no longer discernable and only the
broadest constraints could be placed on $\logg$ from these data.
Even more critical than high photometric precision is 
access to near-ultraviolet (NUV) data,
as can be seen from \fig{twocolour} (c).
In the longer wavelength optical bands
WDs' colours depend primarily on $\teff$,
with minimal separation in $\logg$, or between H and He models. 

However even non-ideal observations of the sort shown in 
\fig{twocolour}~(b) or (c) would allow $\teff$ to be constrained,
provided only that the WD is uambiguously detected 
(\ie, with relative photometric errors of $\la 0.1$)
in at least two passbands that are 
blueward of its Rayleigh--Jeans tail.
In this situation the only other available information 
comes from the flux measurements,
which then fix the solid angle, $\Omega$, subtended by the source.
As 
\begin{equation}
\label{equation:solidangle}
\Omega \propto \frac{R^2}{D^2} \propto \frac{M}{g D^2}
  \propto \frac{M}{10^{\logg} D^2},
\end{equation}
this leads to a well-known degeneracy between distance and radius,
and hence surface gravity,
that is difficult to break with photometric data alone
(\eg, \citealt{Bergeron:1997, Bergeron:2001}).
It is, however, 
possible to use prior knowledge of the Galactic WD population 
-- specifically, that most WDs have $\logg \simeq 8$ --
to break this degeneracy.
\cite{Harris_etal:2006} adopted this approach in their analysis 
of a large sample of photometric SDSS WDs by assuming $\logg = 8$,
although it is undoubtedly overly-prescriptive to 
deny any possibility that a WD's surface gravity deviates from this 
fiducial value.
Nonetheless,
this is more realistic than the alternative
of treating all $\logg$ values as equally likely a priori;
this assumption was made by \cite{Vidrih:2007}, hereafter \vidrih,
in their analysis of a (different) photometric SDSS WD sample.
Both \cite{Harris_etal:2006} and \vidrih\ identified possible
halo WDs, although the existence of a strong parameter 
degeneracy means that the identification of outliers 
on the basis of best-fit parameter estimates is likely to yield 
large numbers of spurious candidates.

The overall implication is that any non-spectroscopic WD parameter estimates
will depend in part on the fitting procedure used,
and that prior information about the Galactic WD population could be as
important as the data that are specific to the source in question.
At least some of these issues can be resolved by adopting a 
more rigorous statistical approach to estimating WD parameters
from photometric and astrometric data. 
This rather broad topic will be addressed in a series of papers,
of which this is the first.
It deals with the more focussed question:
How much can be inferred about a WD from photometric data alone?
After describing the 
Bayesian formalisms for parameter estimation 
(\sect{parameterest})
and model selection (\sect{modelselection}),
the WD population model used as a prior distribution 
is developed in \sect{wdpop}.
After analysing simulated WD observations 
(\sect{simulations}), 
these methods are applied 
to the small samples of candidate 
halo and ultra-cool WDs identified by \vidrih\ in \sect{outliers}.
The conclusions and future directions of this research
are summarised in \sect{conc}.


\section{Parameter estimation}
\label{section:parameterest}

Given photometry of a WD
[\ie, measured apparent magnitudes, $\vect{\data} = 
  (\mappobs_1, \mappobs_2, \ldots, \mappobs_{N_\band})$,
in each of $\nband$ bands],
what constraints can be placed on its properties
[\ie, the model parameters $\parameters = (\loggpar, \teff, \dist)$],
assuming its dominant atmospheric element (H or He) is known?
This is a classic parameter estimation problem for which 
several approaches exist; 
central to all of them is the sampling distribution, or likelihood,
which gives the probability of observing the data
$\data$ given a model $\parameters$.
For the high signal--to--noise ratio measurements\footnote{In reality
the photometric noise includes
both a Poisson contribution due to the finite photon counts and 
a background uncertainty that is usually taken to be Gaussian in flux units.
However, the WD samples considered here have photometric data
accurate to a few \percent, sufficient
to adopt the approximation that 
the noise is additive and Gaussian in magnitude units.
It is further assumed that there is no covariance between 
bands: although there are measurable inter-band noise
correlations in data from a single SDSS scan \citep{Scranton:2005},
this effect is far weaker when using multi-epoch data.}
considered here,
the likelihood has the form 
\begin{eqnarray}
\label{equation:likelihood}
\likelihood(\data|\parameters) 
  & = &
  \likelihood\left(\mappobs_1, \mappobs_2, \ldots, \mappobs_{N_\band}
    | \loggpar, \teff, \dist \right) \nonumber \\
  & = & \prod_{\band = 1}^{N_\band}
    \frac{1}{(2 \pi)^{1/2} \Delta \mappobs_\band} 
    \exp\left(- \chi^2_\band / 2 \right),
\end{eqnarray}
where
\begin{equation}
\label{equation:chisq}
\chi^2_\band = 
\left\{ 
  \frac{ 
    \mappobs_\band - 
  \left[ \mabs_\band(\loggpar, \teff) + 5 \log (\dist / 10 {\unit{pc}}) \right]
  }
  {\Delta \mappobs_\band} \right\}^2,
\end{equation}
$\Delta \mappobs_\band$ is the estimated noise in the $b$th band
and $\mabs_\band(\loggpar,\teff)$ is given 
by interpolating between a grid of WD models.
The effects of the interstellar medium (ISM) are ignored here, 
but could be incorporated into \eq{chisq}
by adding a band-dependent extinction term inside the square brackets 
to account for the extra dimming.

It is at this point that the various methods of parameter
estimation diverge.
Historically, the most common approach has been to use one of several 
\bestfit\ estimators (\sect{bestfit}), 
although it is possible to obtain more reliable parameter values 
and uncertainties using Bayesian inference (\sect{bayes}).

\subsection{\Bestfit\ parameter estimation}
\label{section:bestfit}

An intuitively appealing and commonly used method of parameter
estimation is to find the model, $\parametersobs$, 
that is the best fit to the data.
This can be found in several different ways, including 
maximizing the likelihood,
minimizing $\chi^2$,
or by a least squares approach,
all of which are equivalent for the simple sampling 
distribution given in \eq{likelihood}. 
The large WD samples identified by 
\cite{Kilic:2006}, \cite{Harris_etal:2006} and \vidrih\ 
were analysed using variants of this approach,
although \cite{Harris_etal:2006} restricted their investigation to 
models with $\logg = 8$ and \vidrih\ examined only models 
on Bergeron's grid of pre-computed values.
\cite{Harris_etal:2006} do not quote uncertainties,
and a heuristic procedure to obtain uncertainties is used in \vidrih, 
whereby 
the best and second-best fitting grid points are compared.
Given the fundamental reduction in precision attainable when using only
photometric data to constrain the properties of WDs,
it is clearly important to investigate this issue more thoroughly.

\subsection{Inferential parameter estimation}
\label{section:bayes}

In order to obtain the tightest possible constraints on model 
parameters all the available information must be utilised,
which implies that any 
prior knowledge of the model parameters must be included
(which is also necessary for self-consistency; \eg, \citealt{Jaynes:2003}).
Applying Bayes's theorem to the data, $\data$, 
yields the posterior probability distribution 
\begin{equation}
\label{equation:posterior}
\posterior(\parameters | \data)
  = \frac{\prior(\parameters)
    \likelihood(\data | \parameters)}
    {\evidence(\data)},
\end{equation}
where 
$\prior(\parameters)$
is the prior probability distribution of the parameters,
$\likelihood(\data | \parameters)$
is the likelihood, given in \eq{likelihood},
and the evidence is the integral over the 
$\nparam$ model parameters,
\begin{equation}
\label{equation:evidence}
\evidence(\data)
  = \int \prior(\parameters)
    \likelihood(\data | \parameters) 
    \, \diff \parameter_1 \, \diff \parameter_2 \, 
    \ldots \, \diff \parameter_{\nparam}.
\end{equation}
The evidence plays an important role in model selection
(\sect{modelselection}),
but in parameter estimation serves only to normalise the posterior, 
and so it is often more convenient to work with the unnormalised posterior,
defined most simply as 
\begin{equation}
\posterior^\prime(\parameters | \data)
  = \prior(\parameters)
    \likelihood(\data | \parameters).
\end{equation}

Formally, the entire posterior distribution is the Bayesian answer to 
any parameter estimation problem,
but even though it can be written down directly 
it is not always useful in practice.
Even with just three model parameters, 
a numerical algorithm is required to determine the location 
and extent of the posterior peak --
and the peak is not necessarily well-defined in the presence of degeneracies.
Instead, it is far more efficient to explore the parameter
space 
using Markov chain Monte Carlo 
(MCMC, \eg, \citealt{Neal:1993,Gilks_etal:1999}) techniques.
Many different MCMC algorithms exist,
although all have the same aim: to generate a sample of model parameter
values, $\{ \parameters_1, \parameters_2, \parameters_3, \ldots \}$,
drawn from the posterior distribution.
For the problem at hand, a fairly simple implementation of the 
Metropolis--Hastings algorithm (\citealt{Metropolis:1953,Hastings:1970}) 
is sufficient,
in which the posterior is sampled by a guided random walk that 
tends to explore the high posterior regions whilst
occasionally exploring the low probability tails of the distribution.
The samples generated from the first stage of this process
(burn in) must be discarded, as they are determined partly by 
the starting value, but eventually the statistical properties of
the chain will depend on $\posterior^\prime(\parameters | \data)$ alone.
Whilst there are only heuristic tests to identify these
two phases of the MCMC process (\eg, \citealt{Gelman:1992}) 
a brute force approach can be used to ensure that the final
chain is well-mixed.

A Markov chain of samples from the posterior has a number of 
desirable properties, one of the most fundamental being that 
it already samples all of the possible sub-sets of model parameters,
a task that otherwise would require a multi-dimensional 
marginalisation integral over the other parameters.
Another advantage of the Monte Carlo approach is that the distribution
of not only the independent model parameters 
(\ie, $\loggpar$, $\teff$ and $\dist$ here) but of any derived 
parameters (\eg, mass, $M$, or radius, $R$) can be obtained simply
by calculating the value of interest for each sample drawn from the posterior.

In many situations 
-- in particular that considered here, 
in which chains are generated for each WD in a large sample -- 
it is impractical to store the full chains.
If the 
posterior is sufficiently localised then it is 
generally more useful to obtain a fiducial 
parameter estimate, $\parametersobs$, and uncertainty, 
$\Delta \parametersobs$, from the chain. 
As the full Bayesian answer to any parameter estimation problem is 
the entire posterior distribution, there can be no single correct
approach; but several algorithms give reasonable
approximations to single-peaked posteriors.
The simplest is to take moments of the distribution, 
thus identifying the mean of the distribution as
the fiducial estimate $\parametersobs$ and using the variance
to obtain $\Delta \parametersobs$.  
Whilst this approach is appealingly straightforward, 
it has the disadvantage of being sensitive to outliers 
(which can occasionally be extreme, due to the stochastic nature of the
MCMC algorithm).
A more robust method of generating parameter estimates and 
uncertainties is to use the median value in each model parameter
to construct $\parametersobs$ and then integrate the 
marginalised probability outwards in 
each direction to enclose the desired fraction of the 
posterior.
This approach is adopted here, with the quoted 
uncertainties enclosing 
the central 68 \percent\ of the marginalised probability
in the parameter of interest.


\section{Model selection}
\label{section:modelselection}

\sect{parameterest} addressed the question of how to estimate
the parameters of a WD given its dominant atmospheric element,
but the same data can also be used to determine
the likely atmospheric composition 
(or even whether the object is a WD at all, 
although this possibility is not explored here).
The conventional method of model comparison (\sect{bestfitmodel})
uses only the quality of the best fit under the different hypotheses,
whereas the Bayesian approach marginalises over the model parameters
and 
also includes a prior on each model (\sect{bayesmodel}).

\subsection{Best-fit model selection}
\label{section:bestfitmodel}

The most common approach to model comparison is to use the 
\bestfit\ statistics described in \sect{bestfit} to
characterise how well the models fit the data.
In the simple case under consideration here 
(\ie, in which the two models have the same parameters)
the resultant model comparison statistic typically reduces to
the likelihood ratio, 
\begin{equation}
\frac{\likelihood_{\rmn{max,\hydrogen}}}{\likelihood_{\rmn{max,\helium}}}
  = \exp\left[ - \frac{1}{2} 
    (\chi^2_{\rmn{min,\hydrogen}} - \chi^2_{\rmn{min,\helium}}) \right],
\end{equation}
where the right hand expression assumes the likelihood has the 
form of \eq{likelihood}. 
In some simple situations 
this is a sufficient statistic 
for model comparison, but the fact that it only encodes 
information about a single point in parameter space 
means that it is of limited utility in general.

\subsection{Inferential model selection}
\label{section:bayesmodel}

The central relationship in Bayesian model selection 
(\eg, \citealt{Jaynes:2003})
is, as in \sect{bayes}, Bayes's theorem, 
this time applied to the discrete 
hypotheses that the WD atmosphere is predominantly H or He.
Given data $\data$ and the model parameters $\parameters$
(which, in this case, are the same for both hypotheses),
the probability that the WD atmosphere is H is 
\begin{equation}
\label{equation:probh}
\prob_\hydrogen(\data)
  = \frac{\prior_\hydrogen
    \evidence_\hydrogen(\data)}
  {\prior_\hydrogen
    \evidence_\hydrogen(\data)
  + (1 - \prior_\hydrogen)
  \evidence_\helium(\data)},
\end{equation}
where
$\prior_\hydrogen$
is the (prior) probability that a WD has a H atmosphere,
the substitution 
$\prior_\helium = 1 - \prior_\hydrogen$ 
encodes the assumption that only H and He
atmosphere WDs are considered here,
and 
$\evidence_\hydrogen(\data)$ and 
$\evidence_\helium(\data)$ are,
as defined in \eq{evidence},
the evidence values.

Whilst the evidence completely characterises how consistent the 
data are with each model under consideration, 
it is worth emphasising that the numerical value itself is, 
like the likelihood, not particularly meaningful in isolation.
It is more useful to scale the evidence relative to its maximum possible
value (which is a function of the precision of the data, 
as well as the model and priors), but even then a low relative 
value can have two distinct explanations. 
The most intuitive is simply that 
the data are inconsistent with the entire family of models and that
that any resultant fit is inappropriate;
but in most situations there is also the possibility that
the data are a good fit to a particular subset of models which are,
a priori, unlikely.  


\section{The white dwarf population}
\label{section:wdpop}

In the Bayesian approach to parameter estimation and model 
selection described in 
\sects{parameterest} and \ref{section:modelselection},
respectively, inferences 
are obtained by combining the available data specific to a given WD
with information about the Galaxy's WD population.
If the observations of the source in question are
not particularly informative then the prior information on
the WD population can be critical, 
possibly breaking the otherwise troublesome parameter degeneracies
discussed in \sect{intro}.
More formally,
the WD prior
$\prior(\loggpar, \teff, \dist, l, b)$
gives the probability that a 
WD has a surface gravity of $\loggpar$, 
an effective temperature of $\teff$,
and a given Galactic position
(specified by its distance, $\dist$ and Galactic coordinates, $l$ and $b$),
in the absence of any data specific to the object in question.
As such, it is appropriate to use 
$\prior(\loggpar, \teff, \dist, l, b)$
in the analysis of any WD,
irrespective of how it was discovered 
(\eg, whether it was identified in a flux-limited, volume-limited,
or more heterogeneous sample). 

In principle, 
$\prior(\loggpar, \teff, \dist, l, b)$
should encode a complete model of the Galaxy's WD population, 
giving the relative numbers of WDs with different intrinsic properties
and positions.
It should also include the different 
age and temperature distributions of WDs drawn from the
various sub-populations (thin disk; thick disk; halo; \etc).
In practice, however, 
$\prior(\loggpar, \teff, \dist, l, b)$
has to be inferred from existing surveys and models,
and combining the available information to obtain
$\prior(\loggpar, \teff, \dist, l, b)$ is non-trivial.
This higher-level problem of inferring the properties of the 
Galaxy's WD population is not attempted here,
and instead a simpler, more heuristic, approach is adopted 
to obtain a useful analytic WD parameter prior.

The first simplification is to reduce the number of model 
parameters by omitting $l$ and $b$.
This is justified by the fact that 
the angular position of any detected WD is measured 
to a much greater accuracy than the scales on which 
$\prior(\loggpar, \teff, \dist, l, b)$ varies appreciably.
In statistical terms the astrometric data are 
considerably more informative than the prior; 
in more astronomical terms, the positions of almost all optical sources are 
known to better than an arcsec, whereas no Galactic model would 
have meaningful structure on such small angular scales.
However this does not imply that the reduced prior,
$\prior(\loggpar, \teff, \dist)$, is independent of 
$l$ and $b$, a point explored further in \sect{wdgal}.

A further plausible simplification is that
the reduced prior is separable, so that 
\begin{equation}
\label{equation:prior}
\prior(\loggpar, \teff, \dist) 
  = \prior(\loggpar) \prior(\teff) \prior(\dist).
\end{equation}
The priors on surface gravity and temperature are 
given in \sect{wdintr};
the (position-dependent) distance prior is 
given in \sect{wdgal}.

\subsection{Intrinsic properties}
\label{section:wdintr}

The most precise and unambiguous constraints on the 
intrinsic properties of local WDs come from 
spectroscopic studies, with the on-going SDSS campaign described by 
\cite{Kleinman:2004} and \cite{Eisenstein_etal:2006} by far the largest 
such project.
The prior distribution of a WD's intrinsic
properties adopted here is based largely on their results.

Probably the most straightforward conclusion from the study of 
\cite{Eisenstein_etal:2006} is that DA WDs (with H atmospheres)
outnumber DB WDs (with He atmospheres) by $\sim 10$ to 1, with the 
fractional contribution 
to the WD population of other spectral classes 
being significantly smaller again.
Thus only H and He models are considered here,
and the prior probability that a WD has a H atmosphere is
taken to be $\prob_\hydrogen = 0.9$.

One of the most important results from the spectroscopic studies of WDs
is that they have only a small range of surface gravities 
(\eg, \citealt{Bergeron:1992, Liebert:2005, Eisenstein_etal:2006}). 
The fits presented by \cite{Eisenstein_etal:2006}
imply that $\logg$ for DA stars is
distributed approximately as 
\begin{equation}
\label{equation:loggprior}
\prior(\loggpar) 
  \propto
  \exp\left[- \frac{1}{2} \left( \frac{\logg - 7.9}{0.1} \right)^2 \right],
\end{equation}
corresponding to a mass range of $\sim (0.6 \pm 0.1)\,\msun$.
There appears to be more scatter for DB than for DA WDs in the results of 
\cite{Eisenstein_etal:2006}, but it is unclear whether this
spread is intrinsic.
The first major analysis of DBs \citep{Beauchamp:1996} implied 
that they have a narrower mass distribution than DAs,
and the more recent results of \cite{Voss:2007} again show a 
relatively narrow range of masses.
There is some ambiguity in the surface gravity distribution
due to both the limitations of the models 
at low $\teff$ (\citealt{Eisenstein_etal:2006} and references within),
and the selection effect that all flux-limited samples 
preferentially include larger, more luminous, WDs with low $\logg$.
Indeed, the local volume-limited WD sample compiled by 
\cite{Holberg:2008} implies a $\logg \simeq 8.1$, 
which is in mild disagreement with the results obtained 
by applying a volume correction to the 
more representative, 
but flux-limited, sample of \cite{Liebert:2005}.
It is also possible that there are more WDs
with $|\logg - 7.9| \ga 0.5$ than implied by \eq{loggprior},
but the tails of the instrinsic surface gravity distribution 
are not well measured.
Given the overall consensus with previous studies,
\eq{loggprior} is adopted here as the informative 
WD surface gravity prior for both DB and DA WDs.


The present day temperature distribution of WDs is
determined by a combination of their cooling properties 
and the star formation history of the Galaxy,
as the lower temperature white dwarfs have a wide variety of ages.
The temperature distribution of the hottest WDs is 
simpler because they cool very rapidly -- 
at a rate of $\dtdt \propto \teff^{7/2}$ 
according to the models of \cite{Mestel:1952}, 
although the numerical cooling curves of the Fontaine--Bergeron models 
are more complex.
The implication is that 
any WD observed with $\teff \ga 3 \times 10^4\unit{K}$ 
formed within the last $\sim 10^8 \unit{yr}$,
much shorter than the timescale over which the stellar population
of the Galaxy evolves.
Assuming that all WDs are formed with the same initial $\teff$,
these simple arguments imply that the number density 
of the hottest, youngest WDs scales as
$\diff n_{\rm{WD}} /\diff \teff \propto \teff^{-7/2}$.

It is non-trivial to estimate the WD temperature distribution 
empirically, 
even with, \eg, the large WD sample of \cite{Eisenstein_etal:2006}, as
the selection effects 
mentioned above in the context of the 
surface gravity distribution are even stronger for temperature
due to the $L \propto \teff^4$ luminosity dependence.
A standard $1/V_{\rmn{max}}$ estimator 
is only valid for a population of objects with uniform
spatial distribution, 
whereas most WDs with 
$\teff \ga 5 \times 10^4 \unit{K}$
would be detectable by SDSS at distances of 
$\dist \ga  5 \times 10^3\unit{pc}$.
The temperature distribution of all but the coolest WDs can 
only be inferred in the context of 
a full Galactic model (\cf\ \fig{distprior}).

Fortunately, optical and near-UV photometry is generally sufficient to 
unambiguously constrain a WD's effective temperature,
at least up to $\teff \simeq \teffminrj$, 
so the parameter estimation results described in \sect{parameterest}
should not be greatly affected by this uncertainty in the prior.
Above this temperature, these bands cover only the Rayleigh--Jeans
tail of a WD's spectrum, and the measured colours can only provide
a lower limit on the effective temperature of $\teff \ga \teffminrj$.
In a second fortunate coincidence it is in just this 
regime that the temperature distribution inferred from
the above cooling argument is likely be reasonably accurate,
the lower limit on $\teff$ implied by the data being complemented 
by the sharp $\teff^{-7/2}$ dependence of the prior.
One of the eventual aims of this research is to apply the statistical 
approach described in \sect{bayes} to constraining the properties
of the WD population, but for the moment the temperature prior 
is taken to be
\begin{equation}
\label{equation:teffprior}
\pi(\teff) \propto \teff^{-7/2}.
\end{equation}


\subsection{Galactic distribution}
\label{section:wdgal}

\begin{figure*}
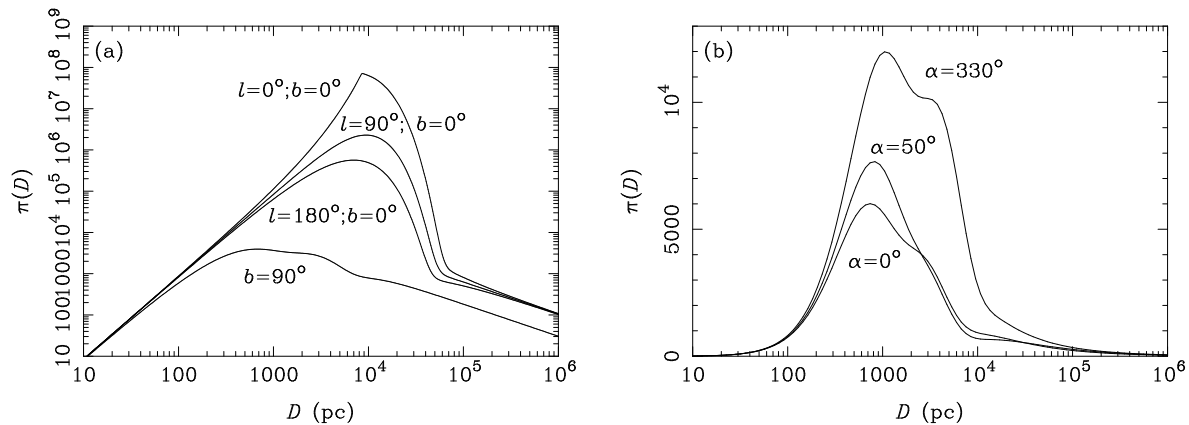

\includegraphics[width=\figurewidth]{\figdir distprior.eps}
\includegraphics[width=\figurewidth]{\figdir distprior_stripe82.eps}
\caption{The prior distributions of $\dist$ that are used here,
  based on the Galactic model of Juri\'{c} \etal\ (2008),
  and described in \sect{wdpop}.
  The prior along four fiducial lines--of--sight is shown in (a).
  The prior along the lines--of--sight to the middle and both ends
  of SDSS Stripe~82 (from which the \vidrih\ sample is drawn) is shown in (b).
  The different lines--of--sight
  are labelled: by Galactic longitude, $l$ and latitude, $b$, in (a);
  and by right ascension (as Stripe~82 is at declination $\delta = 0$) in (b).
  Note that the vertical axes differ,
  and that the distributions are not normalised, instead being
  matched to have the same low-$\dist$ dependence.}
\label{figure:distprior}
\end{figure*}

As WDs are the evolutionary endpoints of MS stars it
is expected that they have a similar distribution in Galactic 
position and velocity, mostly
being located in -- and rotating around the Galactic Centre with --
the stellar disk(s).
This is demonstrably the case for most known WDs, 
although the fact that WDs are so long-lived leads to the intriguing 
possibility of a significant relic population of cool, old, halo WDs 
(\eg, \citealt{Chabrier:1999,Alcock_etal:2000,Hodgkin:2000,Ibata:2000}).
There have been claims that such a population has been detected
(\citealt{Oppenheimer:2001a,Oppenheimer:2001b}), although several studies 
have disputed the inferred space density of this population 
(\citealt{Salim:2004, Reid:2005} and references therein).

The spatial distribution of WDs adopted here includes both 
thin and thick stellar disks, along with the halo, and is based on
the Galactic model described by \cite{Juric:2008}.
In cylindrical Galactic coordinates $R$, $Z$ and $\Phi$,
the number density of stars per unit volume is taken to scale as 
\[
n_{\rmn{MS}}(R, Z, \Phi)
\]
\begin{eqnarray}
\label{equation:rhogal}
\mbox{}
  & \propto & 
    \left[\left[ 
    \exp\left( - \frac{R}{L_\thin} \right) 
    \exp\left( - \frac{|Z|}{H_\thin} \right) 
    \right. \right.
    \nonumber
    \\
  & & +  
    f_\thick \exp\left( - \frac{R}{L_\thick} \right) 
    \exp\left(- \frac{|Z|}{H_\thick} \right)
    \nonumber \\
  & & + 
    \left. \left.
    f_\halo \left\{
      \frac{[R^2 + (Z / q_\halo)^2 + R_\core^2]^{1/2}}{L_\halo}
      \right\}^{- \eta_\halo}
    \right] \right], 
\end{eqnarray}
where 
$f_\thick = 0.06$,
$f_\halo = 6\times10^{-5}$,
$L_\thin = L_\thick = 3500 \unit{pc}$,
$L_\halo = 8500 \unit{pc}$,
$R_\core = 1000 \unit{pc}$,
$H_\thin = 260 \unit{pc}$,
$H_\thick = 1000 \unit{pc}$,
$q_\halo = 0.64$
and
$\eta_\halo = 2$.
This density can then be expressed in terms of 
heliocentric Galactic coordinates, 
defined implicitly by 
\[
R 
  (\dist, l, b) 
  = 
  \left[
  \dist^2 \cos^2(b) - 2 R_\Sun \dist \cos(b) \cos(l) + R_\Sun^2
  \right]^{1/2},
\]
\begin{equation}
\label{equation:coords}
Z 
  (\dist, l, b) 
  = 
  \dist \sin(b), 
\end{equation}
\[
\Phi 
  (\dist, l, b)
  =  
  \arctan \left[ 
  \dist \cos(b) \cos(l) - R_\Sun, \dist \cos(b) \sin(l)
  \right], 
\]
where 
$R_\Sun = 8000 \unit{pc}$ is the fiducial distance from the 
Sun to the Galactic centre.
The fact that the Sun is actually `above' the Galactic plane
with $Z_\Sun = (24.2 \pm 1.7) \unit{pc}$ \citep{MaizApellaniz:2001} is 
unimportant here as the stellar density varies appreciably only 
over much greater scales than this.
The distribution in $\dist$, $l$ and $b$ that results from
substituting these definitions into \eq{rhogal} is rather cumbersome,
and so not given explicitly here.
As argued above, the prior distribution in $l$ and $b$ is 
unimportant,
whereas implied distance distribution has the potential to affect
the parameter estimates for a WD, especially given the 
solid angle degeneracy discussed in \sect{intro}.
The distance prior is obtained by multiplying \eq{rhogal} by the 
volume element ($\propto \dist^2$) to give 
\begin{equation}
\label{equation:distprior}
\pi(\dist | l, b) 
  \propto \dist^2 \, n_{\rmn{MS}}\left[
    R 
    (\dist, l, b),
    Z 
    (\dist, l, b),
    \Phi 
    (\dist, l, b) 
  \right],
\end{equation}
which is shown in in \fig{distprior} for several
values of $l$ and $b$.

The low-$\dist$ behaviour is determined by the volume element;
beyond $\sim 100\unit{pc}$ the Galactic direction is more important,
with the prior rising steeply towards the Galactic centre,
but dropping off sharply when looking out of the plane.
In all cases the high-$\dist$ tail is dominated by the halo
component, which is possibly an underestimate for WDs,
but which should also fall off faster than this power-law form
given in \eq{rhogal} for $\dist \ga 10^5 \unit{pc}$.
A more realistic model would include a sharper fall off in the halo
density, but this is irrelevant for the relatively shallow surveys 
(which can detect WDs only out to $\dist \simeq 10^4 \unit{pc}$)
considered here.

One immediate, if somewhat counter-intuitive, implication of 
the directional dependence of the distance prior is that 
the parameters inferred for two WDs 
with identical photometry could be quite different if they are 
on well separated lines--of--sight.
Consider, for example, a very faint WD with measured colours that,
if it is assumed to have $\logg \simeq 8$,
imply a distance of $\dist \simeq 10^4 \unit{pc}$.
Observed towards the Galactic centre this would be completely
consistent with prior expectation 
(\cf\ the $l = 0 \unit{deg}$; $b = 0 \unit{deg}$ curve in \fig{distprior} a).
Observed out of the Galactic plane this would be very surprising,
as there are expected to be very few WDs at this distance 
(which is much greater than the disk scale heights) 
above or below the Galactic plane,
and it would be much more likely
that the object is a closer, and correspondingly more massive, 
WD with $\logg \simeq 9$.
Most WDs identified in the current generation of wide-field 
surveys are closer than this (\ie, local) and so the variation in the 
Galactic density field is less relevant, but it still has some 
influence in identifying potential halo WDs. 

\subsection{Discussion}

Whilst the prior given by combining 
\eqs{loggprior}, (\ref{equation:teffprior}) and (\ref{equation:distprior})
is a reasonable approximation to the Galaxy's WD population,
it is incomplete in several ways.
The separation of the intrinsic distributions from the 
positional distribution probably leads to the numbers of 
thick disk and halo WDs being underestimated 
(as these stellar populations are older than that of the thin disk).
This model is also equivalent to assuming that the different
Galactic WD populations have the same age and temperature
distributions, which is probably unrealistic, even though there
is considerable uncertainty in this area (\eg, \citealt{Hansen:2001}).
However these approximations are only important here to the degree that
they affect the eventual WD parameter estimates.
For photometric data of the quality considered 
in \sect{simulations} and \ref{section:outliers} the 
most important role of the prior is to encode the 
fiducial surface gravity value and the
relative numbers of thin disk, thick disk and halo WDs,
for which \eqs{loggprior} and (\ref{equation:distprior}) suffice.

\section{Simulations}
\label{section:simulations}

\begin{figure*}
\includegraphics[width=\figurewidth]{\figdir 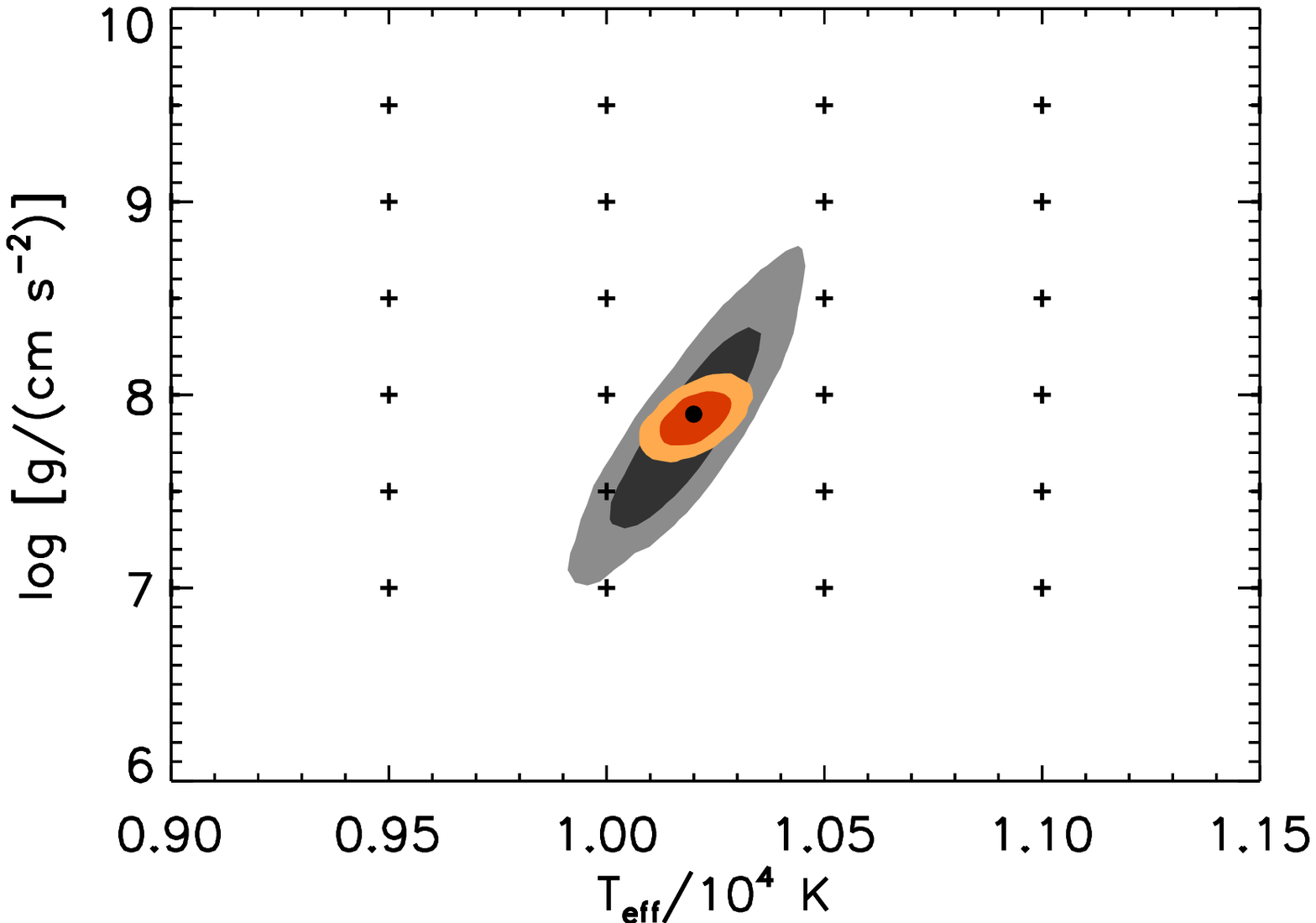}
\includegraphics[width=\figurewidth]{\figdir 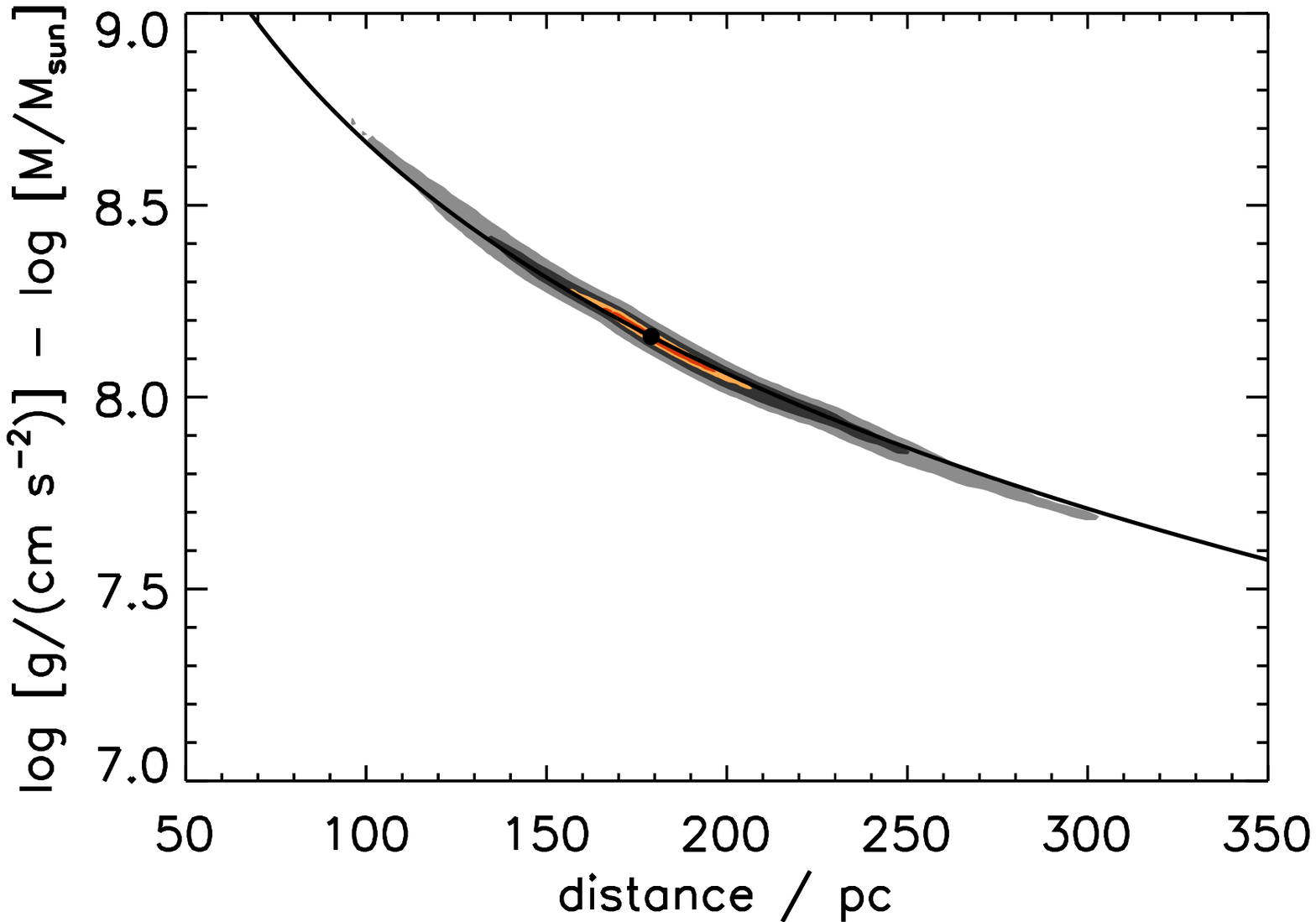}
\caption{Marginalised posterior distributions
  of the model parameters for a simulated WD
  with $\teff = 10200\unit{K}$ and $\logg = 7.9$
  at $\dist = 179\unit{pc}$,
  observed in the $\gprime$, $\rprime$, $\iprime$, $\zprime$, $J$, $H$, $K$
  bands with
  photometric noise of $\Delta\mappobs = 0.01$.
  For both the non-informative prior (grey) and the informative prior
  (orange) the coloured regions enclose 68 \percent\ and 95 \percent\
  of the posterior probability in these two parameters.
  Left: the constraints on the independent model parameters
  $\teff$ and $\logg$ are shown,
  along with the points on the Fontaine--Bergeron model grid (crosses).
  Right: the degeneracy for
  black body models (black line) is highlighted by showing constraints on
  the WD's distance, $\dist$, and radius,
  characterised by $\logg - \log(M / \msun)$.
  In both panels the true model is indicated by the black spot.}
\label{figure:fake1}
\end{figure*}

The implication of the parameter inference and model selection 
methods 
(described in \sects{parameterest} and \ref{section:modelselection})
for WD surveys can be further understood by 
applying them to simulated data.
A number of representative cases are examined
here, although care must be taken in interpreting the 
Bayesian results obtained using the Galactic WD 
prior given in \sect{wdpop},
as the simulated WDs have not been drawn from this
distribution.


In each simulation, a surface gravity, $\loggpar$,
and an effective temperature, $\teff$, 
were chosen from within
the grid of H atmosphere models provided by Bergeron.
This gave absolute AB magnitudes 
in four\footnote{The \uprime\ band was not included in the simulations 
in order to best replicate the analysis of \vidrih\ described in 
\sect{outliers}.}
SDSS bands (\gprime, \rprime, \iprime\ and \zprime)
and three NIR bands ($J$, $H$ and $K$),
which were converted to apparent magnitudes by placing the WD 
at distance $\dist$.
Finally, photometric noise 
(of the same level, $\Delta \hat{m} = 0.01$, in each band, 
as appropriate to the SDSS Stripe~82 data)
was added to give the simulated data
$\hat{\gprime}$, 
$\hat{\rprime}$, 
$\hat{\iprime}$,
$\hat{\zprime}$,
$\hat{J}$,
$\hat{H}$,
and $\hat{K}$.
The data were then analysed to give parameter 
constraints (\sect{simparamest}) and 
also to test the degree to which H and He WDs could be distinguished
(\sect{simmodel}).

\subsection{Parameter estimation}
\label{section:simparamest}

WD parameter estimates were obtained from the simulations in 
several different ways.
Both the 
\bestfit\ model and the \bestfit\ grid value
were calculated as described in \sect{bestfit},
and the full posterior
distributions of the model parameters 
were found using the MCMC algorithm outlined in \sect{bayes}.
In the latter case the analysis was performed both with a non-informative
prior (uniform in the model parameters) and with the full
Galactic WD population prior given in \eq{prior}.

The first case considered is that of a fiducial WD with 
$\logg = 7.9$ and $\teff = 10200\unit{K}$, results for which
are shown in \fig{fake1}.
The most striking feature of this simulation is the 
strong degeneracy apparent in \fig{fake1}~(b),
which follows the form given in \eq{solidangle}.
The temperature is well constrained for both the 
non-informative and informative priors, 
although the weak correlation between $\teff$ and $\logg$ 
means the constraints are better in the latter case.

This example also illustrates an important point about the 
identification of outliers, in particular fast-moving halo WDs.
If this WD was analysed using the grid-based \bestfit\ method
described in \sect{bestfit}, it would be inferred to have
$\loggobs = 7.5$ and $\teffobs = 10000\unit{K}$, simply because this
model grid point is closest to the locus of high likelihood.
The resultant distance estimate would be biased,
high in this case, 
by a factor of $\hat{\dist} / \dist \simeq 1.2$.
The tangential velocity would also be over-estimated,
with the possible result that the WD would be identified as a halo
object if it had a high proper motion.
A more extreme form of this
 effect explains several of the candidate halo WDs analysed in \sect{halo}.
It is also an illustration of the general principle that
it is no more valid to infer that a WD is dynamically interesting
due a large \bestfit\ distance,
than it is to infer that the same star is
intrinsically unusual by restricting its distance through the application of
a prior on its motion.  
The two possibilities should be balanced by 
giving the appropriate (not necessarily equal) weight 
to all the available information.

\begin{figure}
\includegraphics[width=\figurewidth]{\figdir 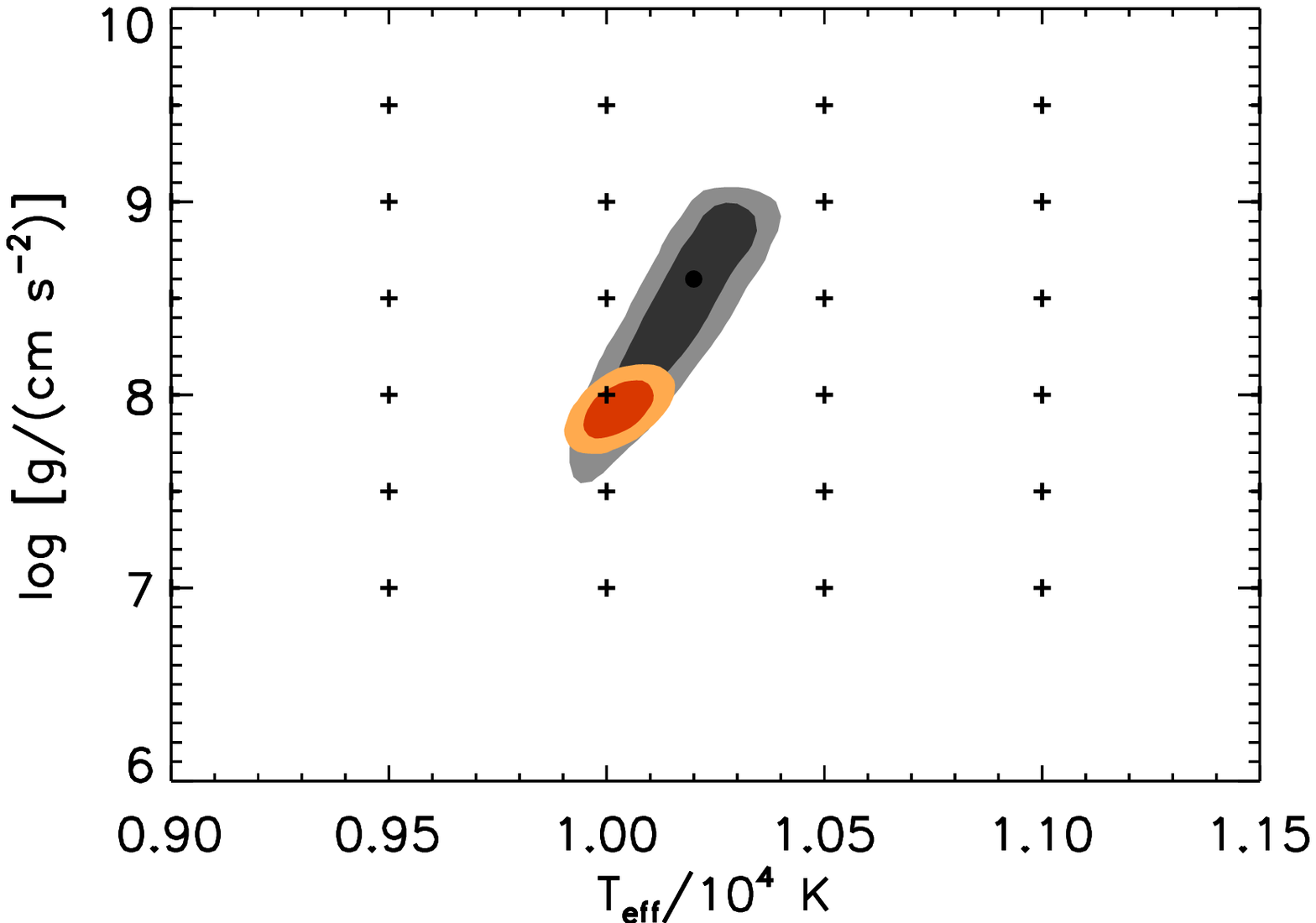}
\caption{Same as \fig{fake1}, but for a simulated WD
  with $\logg = 8.6$ and $\teff = 10200\unit{K}$
  at $\dist = 179 \unit{pc}$.
}
\label{figure:fake2}
\end{figure}

The second example, shown in \fig{fake2}, is of a
$\teff = 10200\unit{K}$ WD with 
an unusually high surface gravity of $\logg = 8.6$.
The results are similar to those for the 
first example, except for the fact that the true model
is well away from the peak of the posterior obtained with the
informative prior.
This comes about as the 
prior based on the Galactic distribution of WDs (with $\logg \simeq 8$)
is inappropriate to the case of an arbitrarily chosen example.  
If a real WD was observed to have these colours and the 
$\logg$ prior given in \eq{loggprior} is correct, this would be
the appropriate inference. 
In that case the small number of WDs 
with true colours that match the data would be out-numbered 
by the fiducial WDs with $\logg \simeq 8$ that were randomly
scattered to these observed values.

\begin{figure}
\includegraphics[width=\figurewidth]{\figdir 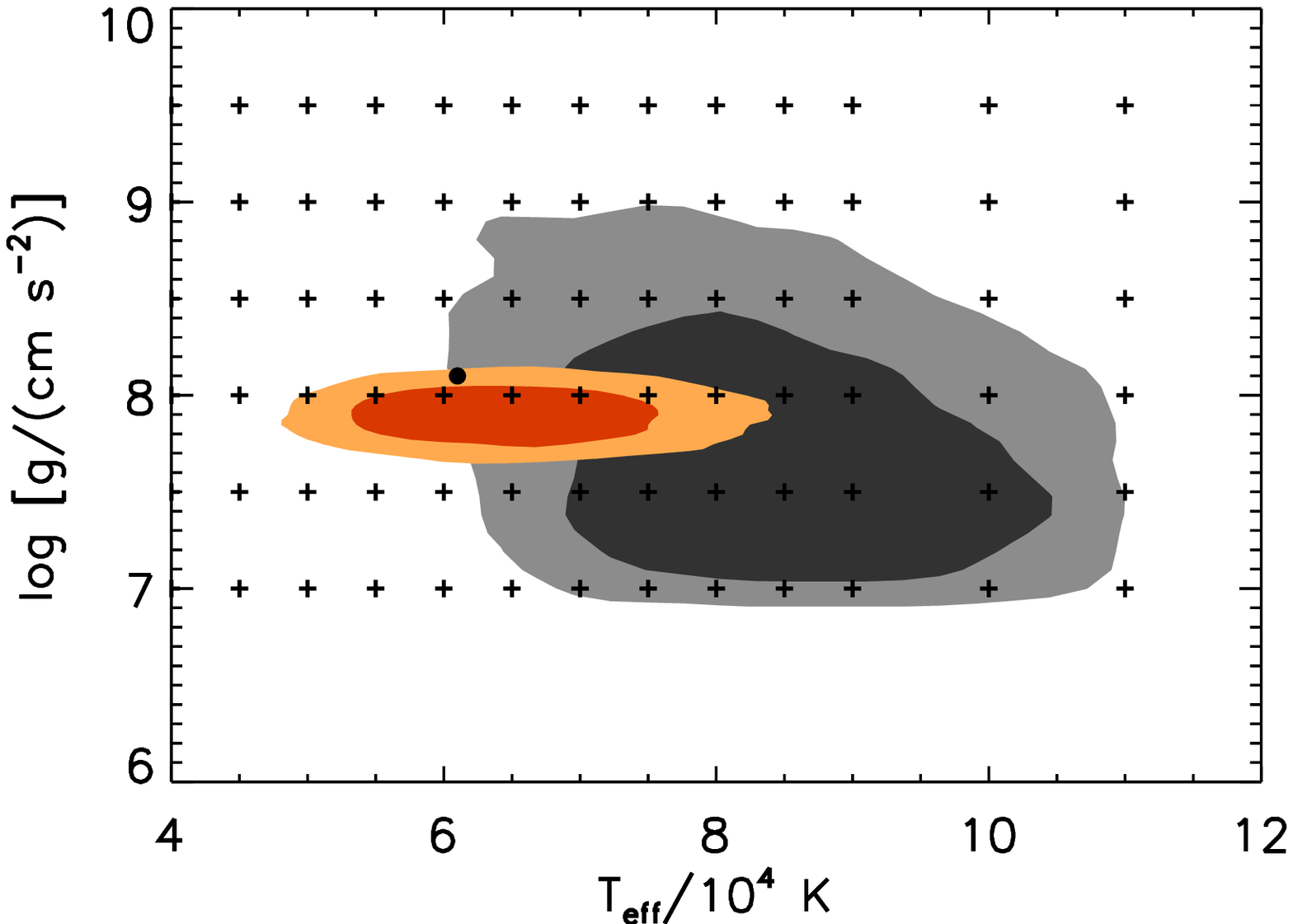}
\caption{Same as \fig{fake1}, but for a simulated WD
  with $\teff = 61000\unit{K}$ and $\logg = 8.1$
  at $\dist = 463 \unit{pc}$.
  }
\label{figure:fake3}
\end{figure}

The results for a hot WD with 
$\teff = 61000\unit{K}$ and $\logg = 8.1$ are shown 
in \fig{fake3}.
The principal difference here
is that the WD's emission peaks at a wavelength of
$\lambda \simeq 0.05\unit{\micron}$, significantly blueward of the SDSS
\gprime\ band, which has a peak response at 
$\lambda \simeq 0.5 \unit{\micron}$.  
As a result it is only the source's
Rayleigh--Jeans tail which is measured, and any temperature
of $\ga \teffminrj$ is consistent with the data.
Both the \bestfit\ analysis and the Bayesian calculation with the
non-informative priors give essentially arbitrary parameter estimates,
although the latter does at least encode the fact that the
uncertainties are large.
However when the WD population model from \sect{wdpop} is
applied, the lower bound on $\teff$ implied by the data
combines with the prior observation that high-$\teff$ WDs are rare
to give a far more precise estimate.
This is a particularly good illustration of the equal role that 
the prior and likelihood have in Bayesian parameter estimation:
in isolation they only provide upper and lower bounds, respectively;
together they combine the available information to 
provide the best possible estimate of the WD's temperature.

\begin{figure}
\includegraphics[width=\figurewidth]{\figdir 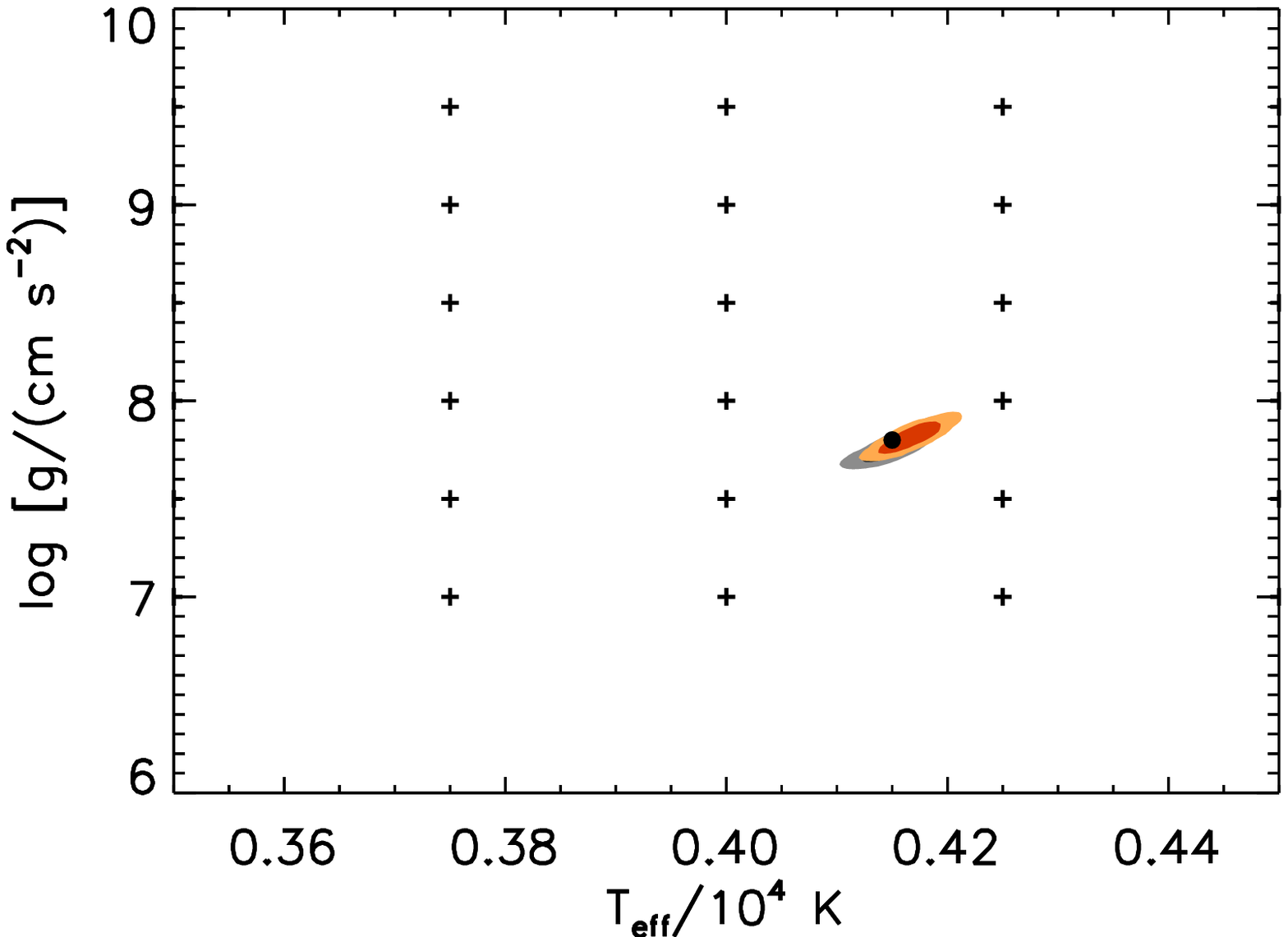}
\caption{Same as \fig{fake1}, but for a simulated WD
  with $\teff = 4150\unit{K}$ and $\logg = 7.8$
  at $\dist = 84 \unit{pc}$.
  }
\label{figure:fake4}
\end{figure}

Results for the final example, a cool WD with 
$\teff = 4150\unit{K}$ and $\logg = 7.8$,
are shown in \fig{fake4}.
At these temperatures the WD colours have a much greater sensitivity
to $\loggpar$, and so the data are almost completely informative,
with the result that the parameter estimates are insensitive to the
nature of the prior, or even to the choice of analysis method.  
The solid angle degeneracy is still apparent,
although the sparsity of the grid (relative to the parameter uncertainties) 
is such that the best fit from a non-interpolative grid-based search
is likely to be outside the formal confidence region in this
temperature range.

\subsection{Model selection}
\label{section:simmodel}

The model selection techniques described in 
\sect{bayesmodel} can be used in a variety of ways;
only one illustrative example is considered here, 
which is to assess the importance of \uprime\ band photometry
to WD parameter estimation.

As discussed in \sect{intro},
the SDSS Stripe~82 data presented by 
\cite{Ivezic:2007} suggests that \percent\ level 
\uprime, \gprime\ and \rprime\ band photometry should be sufficient to both 
determine the surface gravity of H atmosphere WDs to reasonable precision,
and to distinguish between WDs with H and He atmospheres,
at least in the temperature range 
$10000 \unit{K} \la \teff \la 18000 \unit{K}$.
Both these possibilities were explored quantitatively by simulating 
observations of a WD
with $\logg = 8.0$ and $\teff = 13000 \unit{K}$,
using the methodology described above,
but using all the SDSS bands
(\ie, \uprime, \gprime, \rprime, \iprime\ and \zprime).

\begin{figure}
\includegraphics[width=\figurewidth]{\figdir 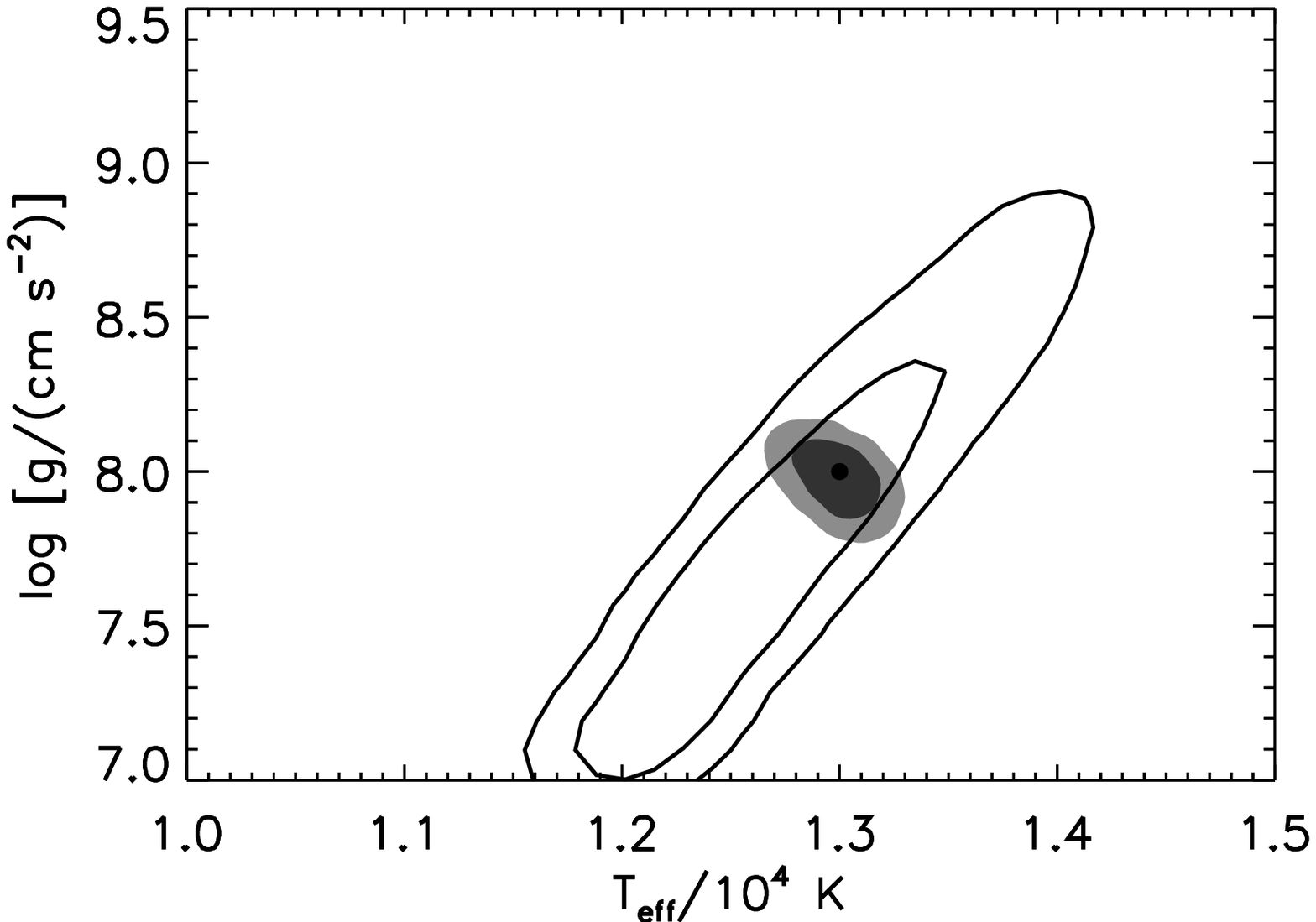}
\caption{Marginalised posterior distributions
  of the model parameters for a simulated WD
  with $\teff = 13000\unit{K}$ and $\logg = 8.0$
  at $\dist = 200\unit{pc}$,
  observed with photometric noise of $\Delta\mappobs = 0.01$ in 
  \uprime, \gprime, \rprime, \iprime\ and \zprime\ (shaded contours)
  and \gprime, \rprime, \iprime\ and \zprime\ only (open contours)
  The contours enclose 68 \percent\ and 95 \percent\
  of the posterior probability (obtained using non-informative priors)
  on the independent model parameters
  $\teff$ and $\logg$. The true model is indicated by the black spot.}
\label{figure:ug}
\end{figure}

The results for parameter estimation, 
obtained with the non-informative priors described
in \sect{wdpop}, are shown in \fig{ug}. 
This confirms that accurate optical and near-UV photometry is sufficient 
to estimate $\logg$ to within $\sim 0.1$, although a second simulation
in only \gprime, \rprime, \iprime\ and \zprime\ also makes it clear that the
$\uprime$ band is critical to doing so.
The loss of the $\uprime$ band also results in increased uncertainties
in $\teff$ as the remaining bands do not probe the Wien 
part of the WD's spectrum as effectively.
These points are particularly relevant to the sample of candidate halo 
WDs analysed by \vidrih\ and reanalysed here in \sect{outliers}.

Given that the separation between the H and He models in \fig{twocolour}
is far greater than that between the various H models 
with $\teff \simeq 13000 \unit{K}$, it is clear that 
photometric data should allow the two classes
of models to be distinguished.
The evidence, defined in \eq{evidence}, was calculated
for H and He models using the informative priors 
defined in \eq{prior}, both with and without \uprime\ band data.  
With \uprime\ measurements the results are as decisive as expected:
$\evidence_\hydrogen \simeq 0.052$ and
$\evidence_\helium \simeq 0.0$,
implying almost no chance that this WD has a He 
atmosphere, with $\posterior_\helium \simeq 10^{-5}$.
However without the \uprime\ band data the situation is not quite as 
clear-cut:
$\evidence_\hydrogen \simeq 7.8$ and
$\evidence_\helium \simeq 1.4$, 
implying $\posterior_\helium \simeq 0.02$.


\section{Outliers}
\label{section:outliers}

One of the main utilities of the 
statistical methods described in \sects{parameterest} and 
\ref{section:modelselection} is the assessment and characterisation of 
putative outliers from the population under consideration.
An example of a problem for which this approach is particularly
well suited is that 
of identifying halo WDs. This subject has provoked much debate
(\eg, \citealt{Oppenheimer:2001a, Oppenheimer:2001b, Reid:2005}),
in part because of the variety of statistical methods used.

One of the largest systematically selected samples of candidate halo 
objects is that generated by \vidrih, who also identified 
a number of candidate ultra-cool, and hence potentially old, WDs.
These samples were extracted from the
SDSS Light--Motion Curve Catalogue (LMCC; \citealt{Bramich_etal:2008}),
which provides data
for $\sim 4 \times 10^6$ sources in SDSS Stripe~82 that 
were 
observed at $\sim 20$ different epochs spanning $\sim 7\unit{yr}$.
As such, the LMCC has improved astrometry and photometry
relative to the main SDSS survey, 
with uncertainties of 
$\Delta \hat{\mu} \simeq 5 \unit{milliarcsec} \unit{yr}^{-1}$
and $\Delta \hat{\rprime} \simeq 0.01$ at $\hat{\rprime} \simeq 20$.
The photometry in the \gprime, \iprime\ and \zprime\ bands is 
of similar quality, although there were fewer good 
\uprime\ band observations, necessitating the exclusion
of the \uprime\ band from the \vidrih\  
analysis, a significant loss in the context of WD studies
(\cf\ \sect{simmodel}).
\vidrih\ were able to produce an exquisite 
reduced proper motion 
(RPM) diagram in which the
WD population was very clearly separated from the MS stars, and 
the resultant sample of 1049 WDs is probably contaminated at no
more than the $\sim 2$ \percent\ level 
(\cf\ \citealt{Kilic:2006, Harris_etal:2006}).

\vidrih\ then cross-matched the WD sample with the
UKIRT Infrared Deep Sky Survey (UKIDSS; \citealt{Lawrence:2006}),
which yielded $Y$, $J$, $H$ and $K$ band photometry 
for about half the sources 
(as the UKIDSS observations in Stripe~82 were only partially complete
at the time).  
The addition of NIR photometry is particularly valuable 
in the analysis of the coolest WDs due to the observed 
depleted flux at these wavelengths \citep{Hodgkin:2000},
although the $Y$ band data were not used by \vidrih\ 
as the Fontaine--Bergeron model grids do not include this filter.

The next stage in \vidrih's analysis was to estimate the WDs' parameters
from the stacked $\gprime$, $\rprime$, $\iprime$ and $\zprime$ band
SDSS photometry 
and, where available, the $J$, $H$ and $K$ UKIDSS measurements.
This was done by finding the best fit from a discrete set of models 
defined by combining the Fontaine--Bergeron grid with a uniform sampling 
(multiples of $5\unit{pc}$) in distance.
This yielded parameter estimates 
$\loggobs$, $\teffobs$ and $\distobs$ 
which, when combined with the observed proper motion, $\hat{\mu}$,
yielded a tangential velocity estimate of $\vtanobs = \distobs \hat{\mu}$.
Uncertainties were characterised by finding the location
of the second-best fitting grid point, 
although these values were not explicitly reported.
\vidrih\ then identified \nhalo\ candidate halo WDs with 
estimated tangential velocities 
of $\vtanobs \geq 160 \kms$,
as well as \nultracool\ candidate ultra-cool WDs with \bestfit\
temperatures of $\teffobs \leq 4000 \unit{K}$.
\vidrih\ took care to point out some of the systematic
effects, including the solid angle degeneracy,
that might affect the candidate selection,
and one of the motivations for this reanalysis is to
quantify some of these biases.


\subsection{Candidate halo white dwarfs}
\label{section:halo}

\begin{table*}
\begin{minipage}{150mm}
\centering
\scriptsize
\caption{Estimated H atmosphere model parameters of the   
  \nhalo\ \vidrih\ candidate halo WDs, 
  obtained using the informative prior defined in \eq{prior}.}
\label{table:haloh}
\begin{tabular}{ccccccc}
\hline
object & $\teff$ & $\logg$ & $M$       & $T$  & $\dist$ & $\vtan$ \\
       & (K)     &         & $(\msun)$ & (yr) & (pc) & (km s$^{-1}$) \\
\hline
SDSS J000244.03$+$010945.8 & $( 1.11 \pm 0.02) \times 10^{ 4} $ & $ 7.88 \pm 0.10 $ & $ 0.55 \pm 0.05 $ & $( 4.04 \pm 0.39) \times 10^{ 8} $ & $   453 \pm 25 $ & $   110 \pm 12 $ \\
SDSS J000557.24$+$001833.2 & $( 9.01 \pm 0.08) \times 10^{ 3} $ & $ 7.89 \pm 0.10 $ & $ 0.54 \pm 0.05 $ & $( 7.06 \pm 0.68) \times 10^{ 8} $ & $   179 \pm 10 $ & $   178 \pm 11 $ \\
SDSS J001306.23$+$005506.4 & $( 1.05 \pm 0.01) \times 10^{ 4} $ & $ 7.88 \pm 0.10 $ & $ 0.54 \pm 0.05 $ & $( 4.63 \pm 0.43) \times 10^{ 8} $ & $   318 \pm 18 $ & $   158 \pm 11 $ \\
SDSS J001518.33$+$010549.2 & $( 1.24 \pm 0.03) \times 10^{ 4} $ & $ 7.86 \pm 0.10 $ & $ 0.54 \pm 0.05 $ & $( 2.91 \pm 0.29) \times 10^{ 8} $ & $   483 \pm 26 $ & $   158 \pm 14 $ \\
SDSS J001838.54$+$005943.5 & $( 1.83 \pm 0.06) \times 10^{ 4} $ & $ 7.87 \pm 0.10 $ & $ 0.56 \pm 0.05 $ & $( 8.24 \pm 1.27) \times 10^{ 7} $ & $   623 \pm 38 $ & $   100 \pm 16 $ \\
SDSS J002951.64$+$005623.6 & $( 1.17 \pm 0.02) \times 10^{ 4} $ & $ 7.88 \pm 0.10 $ & $ 0.55 \pm 0.05 $ & $( 3.50 \pm 0.35) \times 10^{ 8} $ & $   553 \pm 31 $ & $   228 \pm 20 $ \\
SDSS J003054.06$+$001115.6 & $( 7.93 \pm 0.08) \times 10^{ 3} $ & $ 7.89 \pm 0.10 $ & $ 0.54 \pm 0.05 $ & $( 9.93 \pm 1.04) \times 10^{ 8} $ & $   353 \pm 21 $ & $   160 \pm 17 $ \\
SDSS J003730.58$-$001657.8 & $( 8.85 \pm 0.10) \times 10^{ 3} $ & $ 7.89 \pm 0.09 $ & $ 0.54 \pm 0.05 $ & $( 7.46 \pm 0.72) \times 10^{ 8} $ & $   304 \pm 18 $ & $   108 \pm 10 $ \\
SDSS J003813.53$-$000128.8 & $( 1.21 \pm 0.02) \times 10^{ 4} $ & $ 7.87 \pm 0.10 $ & $ 0.55 \pm 0.05 $ & $( 3.17 \pm 0.31) \times 10^{ 8} $ & $   310 \pm 17 $ & $   220 \pm 13 $ \\
SDSS J004214.88$+$001135.7 & $( 1.23 \pm 0.03) \times 10^{ 4} $ & $ 7.88 \pm 0.10 $ & $ 0.55 \pm 0.05 $ & $( 3.09 \pm 0.30) \times 10^{ 8} $ & $   496 \pm 27 $ & $   115 \pm 13 $ \\
SDSS J005906.78$+$001725.2 & $( 1.07 \pm 0.01) \times 10^{ 4} $ & $ 7.88 \pm 0.10 $ & $ 0.54 \pm 0.05 $ & $( 4.48 \pm 0.42) \times 10^{ 8} $ & $   294 \pm 16 $ & $   143 \pm 10 $ \\
SDSS J010129.81$-$003041.7 & $( 1.28 \pm 0.03) \times 10^{ 4} $ & $ 7.88 \pm 0.10 $ & $ 0.55 \pm 0.05 $ & $( 2.76 \pm 0.28) \times 10^{ 8} $ & $   565 \pm 31 $ & $   107 \pm 15 $ \\
SDSS J010207.24$-$003259.7 & $( 1.08 \pm 0.01) \times 10^{ 4} $ & $ 7.89 \pm 0.10 $ & $ 0.55 \pm 0.05 $ & $( 4.40 \pm 0.41) \times 10^{ 8} $ & $   191 \pm 11 $ & $   334 \pm 19 $ \\
SDSS J010225.13$-$005458.4 & $( 1.43 \pm 0.03) \times 10^{ 4} $ & $ 7.88 \pm 0.10 $ & $ 0.56 \pm 0.05 $ & $( 1.97 \pm 0.22) \times 10^{ 8} $ & $   429 \pm 24 $ & $   216 \pm 15 $ \\
SDSS J014247.10$+$005228.4 & $( 1.27 \pm 0.02) \times 10^{ 4} $ & $ 7.87 \pm 0.10 $ & $ 0.55 \pm 0.05 $ & $( 2.81 \pm 0.28) \times 10^{ 8} $ & $   408 \pm 22 $ & $   114 \pm 10 $ \\
SDSS J015227.57$-$002421.1 & $( 9.21 \pm 0.11) \times 10^{ 3} $ & $ 7.89 \pm 0.10 $ & $ 0.54 \pm 0.05 $ & $( 6.69 \pm 0.65) \times 10^{ 8} $ & $   351 \pm 21 $ & $   118 \pm 11 $ \\
SDSS J020241.81$-$005743.0 & $( 7.59 \pm 0.07) \times 10^{ 3} $ & $ 7.89 \pm 0.10 $ & $ 0.54 \pm 0.05 $ & $( 1.10 \pm 0.11) \times 10^{ 9} $ & $   226 \pm 13 $ & $   113 \pm  8 $ \\
SDSS J020729.85$+$000637.6 & $( 5.64 \pm 0.03) \times 10^{ 3} $ & $ 7.88 \pm 0.10 $ & $ 0.53 \pm 0.05 $ & $( 2.50 \pm 0.31) \times 10^{ 9} $ & $   163 \pm  9 $ & $   115 \pm  8 $ \\
SDSS J024529.69$-$004229.8 & $( 1.42 \pm 0.03) \times 10^{ 4} $ & $ 7.87 \pm 0.10 $ & $ 0.55 \pm 0.05 $ & $( 1.98 \pm 0.22) \times 10^{ 8} $ & $   568 \pm 32 $ & $   137 \pm 13 $ \\
SDSS J024837.53$-$003123.9 & $( 8.80 \pm 0.07) \times 10^{ 3} $ & $ 7.89 \pm 0.10 $ & $ 0.54 \pm 0.05 $ & $( 7.54 \pm 0.73) \times 10^{ 8} $ & $   216 \pm 13 $ & $   328 \pm 20 $ \\
SDSS J025325.83$-$002751.5 & $( 1.43 \pm 0.03) \times 10^{ 4} $ & $ 7.80 \pm 0.10 $ & $ 0.52 \pm 0.05 $ & $( 1.77 \pm 0.20) \times 10^{ 8} $ & $   265 \pm 15 $ & $   125 \pm  8 $ \\
SDSS J025531.00$-$005552.8 & $( 1.13 \pm 0.02) \times 10^{ 4} $ & $ 7.88 \pm 0.10 $ & $ 0.55 \pm 0.05 $ & $( 3.88 \pm 0.38) \times 10^{ 8} $ & $   448 \pm 25 $ & $   149 \pm 12 $ \\
SDSS J030433.61$-$002733.2 & $( 6.46 \pm 0.05) \times 10^{ 3} $ & $ 7.89 \pm 0.10 $ & $ 0.54 \pm 0.05 $ & $( 1.65 \pm 0.18) \times 10^{ 9} $ & $   283 \pm 16 $ & $   130 \pm 12 $ \\
SDSS J211928.44$-$002632.9 & $( 1.52 \pm 0.04) \times 10^{ 4} $ & $ 7.87 \pm 0.10 $ & $ 0.55 \pm 0.05 $ & $( 1.60 \pm 0.20) \times 10^{ 8} $ & $   461 \pm 27 $ & $   103 \pm 12 $ \\
SDSS J213641.39$+$010504.9 & $( 1.16 \pm 0.02) \times 10^{ 4} $ & $ 7.88 \pm 0.10 $ & $ 0.55 \pm 0.05 $ & $( 3.55 \pm 0.34) \times 10^{ 8} $ & $   454 \pm 25 $ & $    97 \pm 14 $ \\
SDSS J215138.09$+$003222.3 & $( 7.22 \pm 0.06) \times 10^{ 3} $ & $ 7.89 \pm 0.10 $ & $ 0.54 \pm 0.05 $ & $( 1.26 \pm 0.13) \times 10^{ 9} $ & $   247 \pm 15 $ & $   124 \pm 10 $ \\
SDSS J223808.18$+$003247.9 & $( 6.53 \pm 0.05) \times 10^{ 3} $ & $ 7.88 \pm 0.10 $ & $ 0.53 \pm 0.05 $ & $( 1.59 \pm 0.18) \times 10^{ 9} $ & $   213 \pm 12 $ & $   289 \pm 18 $ \\
SDSS J223815.97$-$011336.9 & $( 7.61 \pm 0.07) \times 10^{ 3} $ & $ 7.88 \pm 0.10 $ & $ 0.53 \pm 0.05 $ & $( 1.08 \pm 0.11) \times 10^{ 9} $ & $   328 \pm 20 $ & $   110 \pm 13 $ \\
SDSS J230534.79$-$010225.2 & $( 1.04 \pm 0.01) \times 10^{ 4} $ & $ 7.89 \pm 0.10 $ & $ 0.55 \pm 0.05 $ & $( 4.88 \pm 0.46) \times 10^{ 8} $ & $   431 \pm 24 $ & $   145 \pm 13 $ \\
SDSS J231626.98$+$004607.0 & $( 4.88 \pm 0.03) \times 10^{ 3} $ & $ 7.89 \pm 0.10 $ & $ 0.52 \pm 0.05 $ & $( 4.99 \pm 0.94) \times 10^{ 9} $ & $   130 \pm  7 $ & $   100 \pm  6 $ \\
SDSS J233227.63$-$010713.8 & $( 1.18 \pm 0.02) \times 10^{ 4} $ & $ 7.88 \pm 0.10 $ & $ 0.55 \pm 0.05 $ & $( 3.47 \pm 0.34) \times 10^{ 8} $ & $   379 \pm 21 $ & $    97 \pm  9 $ \\
SDSS J233817.06$-$005720.1 & $( 9.75 \pm 0.12) \times 10^{ 3} $ & $ 7.88 \pm 0.10 $ & $ 0.54 \pm 0.05 $ & $( 5.75 \pm 0.55) \times 10^{ 8} $ & $   440 \pm 26 $ & $   177 \pm 16 $ \\
SDSS J234110.13$+$003259.5 & $( 1.21 \pm 0.02) \times 10^{ 4} $ & $ 7.87 \pm 0.10 $ & $ 0.54 \pm 0.05 $ & $( 3.18 \pm 0.32) \times 10^{ 8} $ & $   342 \pm 19 $ & $   177 \pm 12 $ \\
SDSS J235138.85$+$002716.9 & $( 7.08 \pm 0.06) \times 10^{ 3} $ & $ 7.89 \pm 0.10 $ & $ 0.54 \pm 0.05 $ & $( 1.32 \pm 0.14) \times 10^{ 9} $ & $   321 \pm 19 $ & $   146 \pm 15 $ \\
\hline
\end{tabular}
\end{minipage}
\end{table*}

\begin{table*}
\begin{minipage}{150mm}
\centering
\scriptsize
\caption{Estimated He atmosphere model parameters of the 
  \nhalo\ \vidrih\ candidate halo WDs,
  obtained using the informative prior defined in \eq{prior}.}
\label{table:halohe}
\begin{tabular}{ccccccc}
\hline
object & $\teff$ & $\logg$ & $M$       & $T$  & $\dist$ & $\vtan$ \\
       & (K)     &         & $(\msun)$ & (yr) & (pc) & (km s$^{-1}$) \\
\hline
SDSS J000244.03$+$010945.8 & $( 1.12 \pm 0.02) \times 10^{ 4} $ & $ 7.89 \pm 0.10 $ & $ 0.53 \pm 0.05 $ & $( 4.29 \pm 0.52) \times 10^{ 8} $ & $   436 \pm 27 $ & $   105 \pm 12 $ \\
SDSS J000557.24$+$001833.2 & $( 8.87 \pm 0.08) \times 10^{ 3} $ & $ 7.89 \pm 0.10 $ & $ 0.53 \pm 0.05 $ & $( 8.02 \pm 0.94) \times 10^{ 8} $ & $   172 \pm 10 $ & $   171 \pm 11 $ \\
SDSS J001306.23$+$005506.4 & $( 1.07 \pm 0.01) \times 10^{ 4} $ & $ 7.89 \pm 0.10 $ & $ 0.53 \pm 0.05 $ & $( 4.92 \pm 0.59) \times 10^{ 8} $ & $   306 \pm 19 $ & $   152 \pm 11 $ \\
SDSS J001518.33$+$010549.2 & $( 1.27 \pm 0.02) \times 10^{ 4} $ & $ 7.89 \pm 0.10 $ & $ 0.54 \pm 0.05 $ & $( 3.09 \pm 0.40) \times 10^{ 8} $ & $   472 \pm 30 $ & $   154 \pm 15 $ \\
SDSS J001838.54$+$005943.5 & $( 1.96 \pm 0.07) \times 10^{ 4} $ & $ 7.90 \pm 0.10 $ & $ 0.55 \pm 0.04 $ & $( 7.42 \pm 1.31) \times 10^{ 7} $ & $   678 \pm 44 $ & $   109 \pm 18 $ \\
SDSS J002951.64$+$005623.6 & $( 1.19 \pm 0.02) \times 10^{ 4} $ & $ 7.89 \pm 0.10 $ & $ 0.54 \pm 0.05 $ & $( 3.69 \pm 0.47) \times 10^{ 8} $ & $   535 \pm 34 $ & $   221 \pm 21 $ \\
SDSS J003054.06$+$001115.6 & $( 7.84 \pm 0.08) \times 10^{ 3} $ & $ 7.90 \pm 0.10 $ & $ 0.53 \pm 0.05 $ & $( 1.11 \pm 0.14) \times 10^{ 9} $ & $   349 \pm 21 $ & $   159 \pm 16 $ \\
SDSS J003730.58$-$001657.8 & $( 8.75 \pm 0.10) \times 10^{ 3} $ & $ 7.89 \pm 0.10 $ & $ 0.53 \pm 0.05 $ & $( 8.34 \pm 0.99) \times 10^{ 8} $ & $   297 \pm 18 $ & $   106 \pm  9 $ \\
SDSS J003813.53$-$000128.8 & $( 1.25 \pm 0.02) \times 10^{ 4} $ & $ 7.89 \pm 0.10 $ & $ 0.53 \pm 0.05 $ & $( 3.18 \pm 0.40) \times 10^{ 8} $ & $   308 \pm 19 $ & $   219 \pm 15 $ \\
SDSS J004214.88$+$001135.7 & $( 1.24 \pm 0.02) \times 10^{ 4} $ & $ 7.89 \pm 0.10 $ & $ 0.53 \pm 0.05 $ & $( 3.25 \pm 0.43) \times 10^{ 8} $ & $   486 \pm 31 $ & $   113 \pm 14 $ \\
SDSS J005906.78$+$001725.2 & $( 1.08 \pm 0.01) \times 10^{ 4} $ & $ 7.89 \pm 0.10 $ & $ 0.53 \pm 0.05 $ & $( 4.73 \pm 0.57) \times 10^{ 8} $ & $   284 \pm 18 $ & $   138 \pm 10 $ \\
SDSS J010129.81$-$003041.7 & $( 1.29 \pm 0.02) \times 10^{ 4} $ & $ 7.89 \pm 0.10 $ & $ 0.54 \pm 0.05 $ & $( 2.91 \pm 0.39) \times 10^{ 8} $ & $   559 \pm 35 $ & $   106 \pm 15 $ \\
SDSS J010207.24$-$003259.7 & $( 1.10 \pm 0.01) \times 10^{ 4} $ & $ 7.89 \pm 0.10 $ & $ 0.53 \pm 0.05 $ & $( 4.54 \pm 0.54) \times 10^{ 8} $ & $   186 \pm 11 $ & $   327 \pm 20 $ \\
SDSS J010225.13$-$005458.4 & $( 1.47 \pm 0.03) \times 10^{ 4} $ & $ 7.89 \pm 0.10 $ & $ 0.54 \pm 0.05 $ & $( 1.97 \pm 0.29) \times 10^{ 8} $ & $   442 \pm 29 $ & $   222 \pm 17 $ \\
SDSS J014247.10$+$005228.4 & $( 1.29 \pm 0.02) \times 10^{ 4} $ & $ 7.89 \pm 0.10 $ & $ 0.54 \pm 0.05 $ & $( 2.91 \pm 0.38) \times 10^{ 8} $ & $   405 \pm 25 $ & $   113 \pm 10 $ \\
SDSS J015227.57$-$002421.1 & $( 9.13 \pm 0.11) \times 10^{ 3} $ & $ 7.89 \pm 0.10 $ & $ 0.53 \pm 0.05 $ & $( 7.46 \pm 0.88) \times 10^{ 8} $ & $   340 \pm 21 $ & $   115 \pm 11 $ \\
SDSS J020241.81$-$005743.0 & $( 7.46 \pm 0.06) \times 10^{ 3} $ & $ 7.89 \pm 0.10 $ & $ 0.52 \pm 0.05 $ & $( 1.25 \pm 0.15) \times 10^{ 9} $ & $   222 \pm 13 $ & $   112 \pm  8 $ \\
SDSS J020729.85$+$000637.6 & $( 5.67 \pm 0.03) \times 10^{ 3} $ & $ 7.93 \pm 0.10 $ & $ 0.54 \pm 0.05 $ & $( 3.06 \pm 0.46) \times 10^{ 9} $ & $   164 \pm 10 $ & $   116 \pm  8 $ \\
SDSS J024529.69$-$004229.8 & $( 1.45 \pm 0.03) \times 10^{ 4} $ & $ 7.90 \pm 0.10 $ & $ 0.54 \pm 0.05 $ & $( 2.08 \pm 0.30) \times 10^{ 8} $ & $   573 \pm 37 $ & $   139 \pm 14 $ \\
SDSS J024837.53$-$003123.9 & $( 8.67 \pm 0.08) \times 10^{ 3} $ & $ 7.90 \pm 0.09 $ & $ 0.53 \pm 0.05 $ & $( 8.67 \pm 1.01) \times 10^{ 8} $ & $   208 \pm 12 $ & $   316 \pm 19 $ \\
SDSS J025325.83$-$002751.5 & $( 1.49 \pm 0.03) \times 10^{ 4} $ & $ 7.89 \pm 0.10 $ & $ 0.54 \pm 0.05 $ & $( 1.88 \pm 0.27) \times 10^{ 8} $ & $   263 \pm 17 $ & $   125 \pm  9 $ \\
SDSS J025531.00$-$005552.8 & $( 1.14 \pm 0.02) \times 10^{ 4} $ & $ 7.89 \pm 0.10 $ & $ 0.53 \pm 0.05 $ & $( 4.09 \pm 0.50) \times 10^{ 8} $ & $   432 \pm 27 $ & $   143 \pm 12 $ \\
SDSS J030433.61$-$002733.2 & $( 6.40 \pm 0.04) \times 10^{ 3} $ & $ 7.91 \pm 0.10 $ & $ 0.53 \pm 0.05 $ & $( 1.88 \pm 0.24) \times 10^{ 9} $ & $   280 \pm 17 $ & $   129 \pm 12 $ \\
SDSS J211928.44$-$002632.9 & $( 1.56 \pm 0.04) \times 10^{ 4} $ & $ 7.90 \pm 0.10 $ & $ 0.55 \pm 0.05 $ & $( 1.64 \pm 0.25) \times 10^{ 8} $ & $   474 \pm 32 $ & $   106 \pm 13 $ \\
SDSS J213641.39$+$010504.9 & $( 1.18 \pm 0.02) \times 10^{ 4} $ & $ 7.89 \pm 0.10 $ & $ 0.53 \pm 0.05 $ & $( 3.75 \pm 0.47) \times 10^{ 8} $ & $   438 \pm 28 $ & $    94 \pm 14 $ \\
SDSS J215138.09$+$003222.3 & $( 7.12 \pm 0.06) \times 10^{ 3} $ & $ 7.90 \pm 0.10 $ & $ 0.53 \pm 0.05 $ & $( 1.41 \pm 0.17) \times 10^{ 9} $ & $   244 \pm 15 $ & $   123 \pm 10 $ \\
SDSS J223808.18$+$003247.9 & $( 6.43 \pm 0.04) \times 10^{ 3} $ & $ 7.93 \pm 0.10 $ & $ 0.54 \pm 0.05 $ & $( 1.89 \pm 0.24) \times 10^{ 9} $ & $   206 \pm 12 $ & $   279 \pm 18 $ \\
SDSS J223815.97$-$011336.9 & $( 7.53 \pm 0.07) \times 10^{ 3} $ & $ 7.89 \pm 0.10 $ & $ 0.52 \pm 0.05 $ & $( 1.21 \pm 0.15) \times 10^{ 9} $ & $   325 \pm 19 $ & $   109 \pm 13 $ \\
SDSS J230534.79$-$010225.2 & $( 1.04 \pm 0.02) \times 10^{ 4} $ & $ 7.89 \pm 0.10 $ & $ 0.53 \pm 0.05 $ & $( 5.27 \pm 0.63) \times 10^{ 8} $ & $   413 \pm 25 $ & $   139 \pm 13 $ \\
SDSS J231626.98$+$004607.0 & $( 5.13 \pm 0.02) \times 10^{ 3} $ & $ 7.90 \pm 0.10 $ & $ 0.52 \pm 0.05 $ & $( 4.59 \pm 0.81) \times 10^{ 9} $ & $   148 \pm  9 $ & $   114 \pm  8 $ \\
SDSS J233227.63$-$010713.8 & $( 1.20 \pm 0.02) \times 10^{ 4} $ & $ 7.89 \pm 0.10 $ & $ 0.53 \pm 0.05 $ & $( 3.58 \pm 0.45) \times 10^{ 8} $ & $   371 \pm 23 $ & $    95 \pm  9 $ \\
SDSS J233817.06$-$005720.1 & $( 9.74 \pm 0.12) \times 10^{ 3} $ & $ 7.89 \pm 0.10 $ & $ 0.53 \pm 0.05 $ & $( 6.30 \pm 0.74) \times 10^{ 8} $ & $   424 \pm 26 $ & $   171 \pm 16 $ \\
SDSS J234110.13$+$003259.5 & $( 1.23 \pm 0.02) \times 10^{ 4} $ & $ 7.89 \pm 0.10 $ & $ 0.53 \pm 0.05 $ & $( 3.30 \pm 0.42) \times 10^{ 8} $ & $   334 \pm 21 $ & $   173 \pm 12 $ \\
SDSS J235138.85$+$002716.9 & $( 7.00 \pm 0.06) \times 10^{ 3} $ & $ 7.89 \pm 0.10 $ & $ 0.52 \pm 0.05 $ & $( 1.46 \pm 0.18) \times 10^{ 9} $ & $   319 \pm 19 $ & $   145 \pm 15 $ \\
\hline
\end{tabular}
\end{minipage}
\end{table*}

\begin{figure}
\includegraphics[width=\figurewidth]{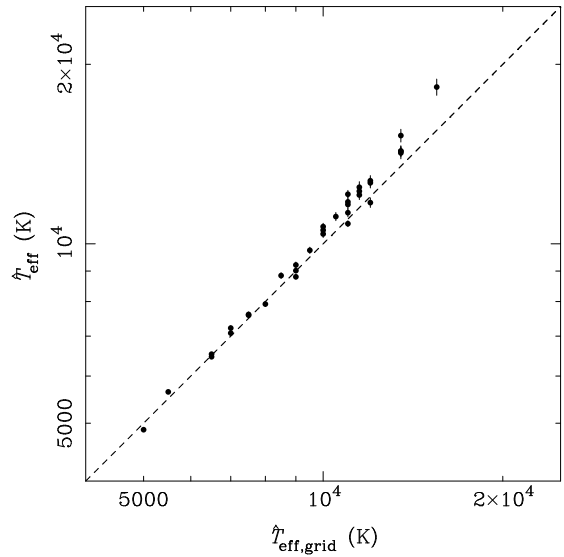}
\caption{The estimated effective temperature, $\teffobs$, of the
  \nhalo\ candidate halo WDs as found by \vidrih,
  and using the Bayesian methods described here 
  (for the H atmosphere models with 
  the Galactic model priors).}
\label{figure:haloteff}
\end{figure}

\begin{figure*}
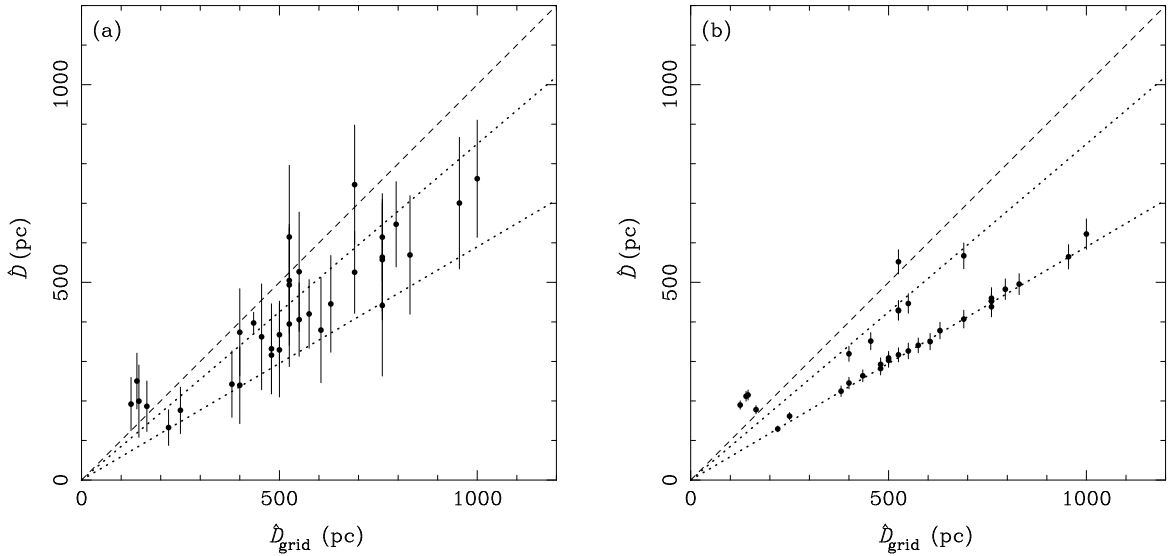

\includegraphics[width=\figurewidth]{\figdir halo_h_flat_dist.eps}
\includegraphics[width=\figurewidth]{\figdir halo_h_eisenstein_dist.eps}
\caption{The estimated distance, $\distobs$, to the
  \nhalo\ candidate halo WDs as calculated by \vidrih\ 
  and obtained using the Bayesian method described here 
  [for H atmosphere models,
  with the non-informative priors in (a) and the Galactic model priors in (b)].
  The two dotted lines show the loci 
  $\distobs = 0.6\, \distobs_{\rmn{grid}}$ and 
  $\distobs = 0.85\, \distobs_{\rmn{grid}}$, expected for 
  WDs with unconstrained surface gravities 
  for which \vidrih\ estimate $\loggobs = 7.0$ and 
  $\loggobs = 7.5$, respectively.}
\label{figure:halodist}
\end{figure*}

\begin{figure}
\includegraphics[width=\figurewidth]{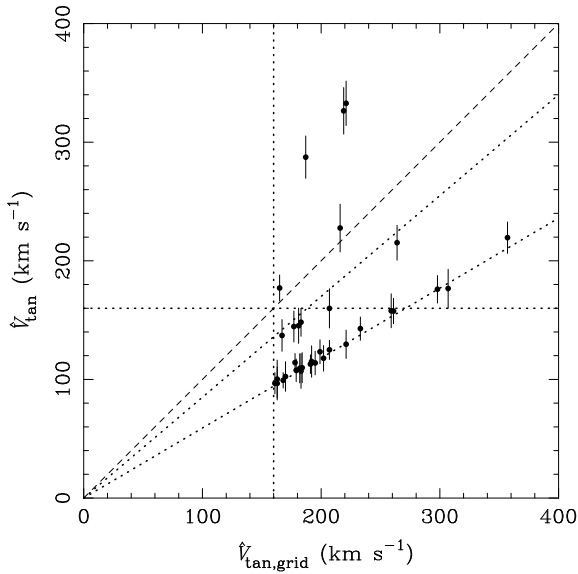}
\caption{The estimated tangential velocity, $\vtanobs$, of the 
  \nhalo\ candidate halo WDs as found by \vidrih\ 
  and obtained using the Bayesian method described here
  (for H atmosphere models,
  with 
  the Galactic model priors).
  The horizontal and vertical dotted lines show the criterion of 
  $\vtan \geq 160\unit{km} \unit{s}^{-1}$ used by \vidrih\ 
  to select their sample of halo candidates. 
  The other two dotted lines show the loci
  $\vtanobs = 0.6\, \vtanobsgrid$ and
  $\vtanobs = 0.85\, \vtanobsgrid$, expected for
  WDs with unconstrained surface gravities
  for which \vidrih\ estimate $\logg = 7.0$ and
  $\logg = 7.5$, respectively.}
\label{figure:halovtan}
\end{figure}

The \nhalo\ candidate halo WDs identified by \vidrih\  
were reanalysed using the Bayesian techniques described in \sects{parameterest}
and \ref{section:modelselection}, giving parameter estimates and 
evidence values for both H and He atmosphere models,
and for both the non-informative and informative priors defined in 
\sect{wdpop}.
The results are presented in \tabls{haloh} and \ref{table:halohe}.
The Bayesian parameter fits are compared to those obtained by 
\vidrih\ in \figs{haloteff}, \ref{figure:halodist} and \ref{figure:halovtan}.
The model comparison results, used to distinguish between the 
H and He models, are presented in \fig{halomodel}.

\fig{haloteff} shows the expected result that the Bayesian 
effective temperature estimates match those of \vidrih.
The estimates are insensitive to the prior, as
the well-measured SDSS (and UKIDSS) colours 
unambiguously constrain $\teff$. 
If the data are sufficiently informative then the details of 
the parameter estimation method are secondary, 
and that is the situation here.
The clear agreement between the two estimates notwithstanding,
the Bayesian values are systematically 
higher; this is an artefact of the much stronger distance bias 
discussed below, which has a small effect here because there is a
slight anti-correlation between $\teffobs$ and $\distobs$ for 
WDs with $\teff \simeq 10^4\unit{K}$.  

The Bayesian distance estimates for the halo candidates
differ considerably
from those given by \vidrih, as can be seen from \fig{halodist}.
For both the non-informative and informative priors the Bayesian 
estimates are lower, for what turn out to be related 
(but distinct) reasons.

In the case of non-informative priors it is because the \vidrih\   
parameter search is performed on a discrete grid of models.  The grid point 
with the lowest $\chi^2$ is unlikely to be the true \bestfit, even if 
it might be close.  In many problems this would merely result in 
a small error in the parameter estimates on a scale comparable to 
the grid spacing.  However, the strong degeneracy in the WD models 
exacerbates this problem, as the locus of models with good fits
can `thread' the model grid, as shown in \figs{fake1}, \ref{figure:fake2} and 
\ref{figure:fake4}.
This explains why the grid-based estimates can differ significantly
from the true best fit distance, but not why they are 
systematically higher.  The bias seen in \fig{halodist}
can be traced back to the 
fact that the \nhalo\ candidate halo WDs were not selected randomly,
but by virtue of having large $\vtanobs$ and hence high $\distobs$.  
Thus WDs for which the error due to the grid-based fitting 
randomly leads to an over-estimate of $\dist$ will preferentially
be selected into a sample of objects with high inferred tangential 
velocities, explaining the observed bias.

The difference between the Bayesian distance estimates obtained 
using the informative prior and those of \vidrih\ are
even greater, as shown in \fig{halodist}~(b).  The selection bias
described above is still present, but 
the strongly peaked $\logg$ prior given in \eq{loggprior} 
is the dominant factor.
With all the models along the degeneracy being comparably good 
fits to the data, the informative prior results in the posterior
being strongly peaked at, in this case, $\logg \simeq 7.9$. 
This effectively breaks the degeneracy, and the tight limits 
on $\logg$ from the prior then translate into a comparably tight
constraint on $\dist$.  The obvious locus of points with 
$\hat{\dist} \simeq 0.6\, \hat{\dist}_{\rmn{grid}}$ are those 
for which the best fit grid point has $\loggobs = 7.0$ and the Bayesian
models have the prior-selected value of $\logg \simeq 7.9$; 
the distance ratio implied by \eq{solidangle} is $0.59$.
There are also a few points with 
$\distobs \simeq 0.85\, \hat{\dist}_{\rmn{grid}}$;
these are the WDs for which the \bestfit\ grid point has $\loggobs = 7.5$.
Whilst it is not necessarily unexpected that a warm, and hence
young, halo WD would have low mass and $\logg$, 
such objects are expected to extremely rare.
A more probable explanation is that these are members of the 
much greater thin disk and thick disk populations
for which the grid-based fitting procedures 
happened to yield the the twin abnormalities of high 
$\distobs$ and low $\loggobs$.
(For a more detailed discussion of this idea see 
\sect{haloindividual}.)
The tightness of the loci in \fig{halodist} is also rather striking, 
and suggests that the informative $\logg$ prior defined in \eq{loggprior}
is overly prescriptive; the true $\logg$ distribution almost certainly has 
broader, non-Gaussian wings, and the resultant distance estimates
probably should lie between the two extremes presented here.

The estimated tangential velocity, $\vtanobs$, is directly related to the 
distance estimates described above by $\vtanobs = \distobs \hat{\mu}$, 
and so it is not surprising 
that \fig{halovtan} exhibits some of the same features as \fig{halodist}.
In particular the Bayesian values of $\vtanobs$ are correspondingly
smaller than those given by \vidrih, and only 
10 of these \nhalo\ sources would satisfy the 
$\vtan \geq 160 \unit{km} \unit{s}^{-1}$ criterion as 
candidate halo WDs if the Bayesian estimates were used.
Just as before, the Gaussian $\logg$ prior is probably 
overly prescriptive, but the essential conclusion,
that the tangential velocities of these objects are lower than
estimated by \vidrih, is fairly robust.
There are also three WDs,
discussed further in \sect{haloindividual},
for which the Bayesian estimate of $\vtan$ is considerably greater than
that quoted by \vidrih.  
They also have the three highest measured 
proper motions of the \nhalo\ WDs in the sample,
which reinforces their identification as halo candidates.
Their measured proper motions also give some information about their 
direction of motion, which could in principle constrain them to have
low $\dist$, most obviously if this implied an unreasonably high
orbital speed along the Solar circle.
However this is not the case for any of these three sources,
and the uncertainty due to the lack of radial velocity information
makes it difficult to say more.

The best way to include this dynamical information would be 
to adopt a velocity prior, in much the same way as the distance 
prior was developed in \sect{wdgal}.
As changing the distance to a WD with fixed physical properties 
and a fixed spatial velocity, $\vel$, would affect both its measured 
photometry and proper motion, it is the only way to correctly include 
the available prior information 
(\ie, on $\loggpar$, $\teff$, $\dist$ and $\vel$) 
and thus utilise all the available data simultaneously
(\cf\ \citealt{Hogg:2005}).

All the above results were obtained under the assumption that the
\vidrih\ halo candidates have 
H atmospheres, but given that 
$\sim 10$ \percent\ of WDs have He atmospheres,
several of the \nhalo\ halo candidates are probably in this category.
This can be assessed objectively by using the model comparison
formalism described in \sect{modelselection}.
Comparing 
$\evidence_\hydrogen$ and $\evidence_\helium$ reveals that
most of the 
\nhalo\ halo candidates are reasonably consistent with either model,
with $|\ln(\evidence_\hydrogen) - \ln(\evidence_\helium)| \la 5$,
although some are decisively better fits to the H models.
However, care must be taken in interpreting the evidence, as its 
numerical value itself is of little meaning in isolation,
especially due to the strong model prior that H atmosphere
WDs are so more common than He WDs.
Adopting the model prior $\prior_\hydrogen = 0.9$ 
(discussed in \sect{wdpop})
allows the posterior model probability,
$\posterior_\hydrogen$,
to be calculated according to \eq{probh}.
These are shown in \fig{halomodel} and compared to 
the commonly used model comparison statistic
$\Delta \chi^2 =
  \chi^2_{{\rmn{min}},\helium} - \chi^2_{{\rmn{min}},\hydrogen}$
(\cf\ \sect{modelselection}).
As expected, this is correlated with $\posterior_\hydrogen$,
but there is a broad scatter in this relation because 
$\Delta \chi^2$ is sensitive only to a single point in parameter
space (\ie, the best fit) whereas $\posterior_\hydrogen$ 
is a properly weighted integral over all the possible models.
There is also a systematic shift due to the strong model prior; 
if it was set to the non-informative value of $\prior_\hydrogen = 0.5$,
all the points would be higher in \fig{halomodel} and 
the more probable hypothesis would be simply that with the 
higher evidence value, as seen from \eq{probh}.


\begin{figure}
\includegraphics[width=\figurewidth]{\figdir halo_modelcomp.eps}
\caption{A comparison of the H and He fits to the \nhalo\ candidate halo WDs.
  The difference in $\chi^2_{\rmn{min}}$ is compared to the 
  Bayesian probability that the WD has a H atmosphere, 
  $\posterior_\hydrogen$, calculated assuming the prior probability
  that any WD has a H atmosphere is $\prior_\hydrogen = 0.9$.
  The dashed horizontal line is located at
  $\posterior_\hydrogen = \posterior_\helium = 0.5$;
  the eight WDs above this line probably have He atmospheres, though in no 
  case is this conclusion decisive. Each star in the subset of the probable 
  He stars for which spectroscopic confirmation exists in the Eisenstein 
  \etal\ (2006) sample has been classified as type DA. }
\label{figure:halomodel}
\end{figure}

In order to investigate these conclusions further, \tabl{halomodel}
gives some of the quantities used in model selection. The best fit
$\logg$ obtained with the uninformative $\logg$ prior is shown
as a proxy for the influence of the informative $\logg$ prior -- the further 
this is from $\logg =7.9$, the stronger the effect of the $\loggpar$ 
prior will be in 
the model selection. The best fit $\chisq$ 
(for the uninformative prior)
is shown as an indicator of the influence of the data.
The evidence values are presented to indicate how consistent the 
data are with each model over the entire model space. If neither
the $\loggpar$ prior nor the data definitively picks out an atmosphere model,
the 10:1 model prior dominates the model selection. Finally,
the probability of the WD having a H atmosphere is listed, 
along with an indication of the driving force behind this conclusion. 

\begin{table}
\begin{minipage}{81mm}
\centering
\scriptsize
\caption{Estimated model parameters of the
  \neisen\ \vidrih\ candidate halo WDs
  with spectroscopic fits presented by Eisenstein \etal\ (2006).
  Three WDs, SDSS~J024837.53$-$003123.9,
  SDSS~J024837.53$-$003123.9 and
  SDSS J234110.13$+$003259.5, were observed twice.}
\label{table:haloeisen}
\begin{tabular}{cccc}
\hline
object & type & $\loggobs$ & $\teffobs$ \\
       &      &         &    (K)   \\
\hline
SDSS J000557.24$+$001833.2 & DZ & $7.78 \pm 0.58$ &   $10110 \pm   50$ \,\,\, \\
SDSS J001306.23$+$005506.4 & DA & $8.13 \pm 0.14$ &   $10890 \pm  170$ \, \\
SDSS J001838.54$+$005943.5 & DA & $7.64 \pm 0.13$ &   $21820 \pm  760$ \, \\
SDSS J003730.58$-$001657.8 & DA & $7.93 \pm 0.38$ & \, $9130 \pm  160$ \, \\
SDSS J003813.53$-$000128.8 & DA & $8.03 \pm 0.11$ &   $12140 \pm  400$ \, \\
SDSS J005906.78$+$001725.2 & DZ & $8.53 \pm 0.38$ &   $11180 \pm  270$ \, \\
SDSS J010207.24$-$003259.7 & DA & $8.24 \pm 0.08$ &   $11050 \pm  100$ \, \\
SDSS J010225.13$-$005458.4 & DA & $8.95 \pm 0.06$ &   $18790 \pm  830$ \, \\
SDSS J014247.10$+$005228.4 & DA & $8.82 \pm 0.11$ &   $14040 \pm  450$ \, \\
SDSS J024837.53$-$003123.9 & DA & $7.81 \pm 0.15$ & \, $9080 \pm   80$ \, \, \\
                           & DA & $8.11 \pm 0.17$ & \, $9500 \pm  110$ \, \\
SDSS J025325.83$-$002751.5 & DA & $7.70 \pm 0.03$ &   $19200 \pm  200$ \, \\
SDSS J025531.00$-$005552.8 & DA & $7.72 \pm 0.20$ &   $12720 \pm  640$ \, \\
                           & DA & $7.74 \pm 0.15$ &   $12830 \pm  440$ \, \\
SDSS J211928.44$-$002632.9 & DA & $7.87 \pm 0.10$ &   $15660 \pm  410$ \, \\
SDSS J215138.09$+$003222.3 & DA & $8.23 \pm 0.45$ & \, $7820 \pm  210$ \, \\
SDSS J233227.63$-$010713.8 & DA & $7.60 \pm 0.20$ &   $14420 \pm 1320$ \\
SDSS J233817.06$-$005720.1 & DA & $8.07 \pm 0.33$ &   $10380 \pm  280$ \, \\
SDSS J234110.13$+$003259.5 & DA & $7.90 \pm 0.08$ &   $13380 \pm  330$ \, \\
                           & DA & $7.80 \pm 0.09$ &   $12830 \pm  280$ \, \\
\hline
\end{tabular}
\end{minipage}
\end{table}

\begin{figure}
\includegraphics[width=\figurewidth]
  {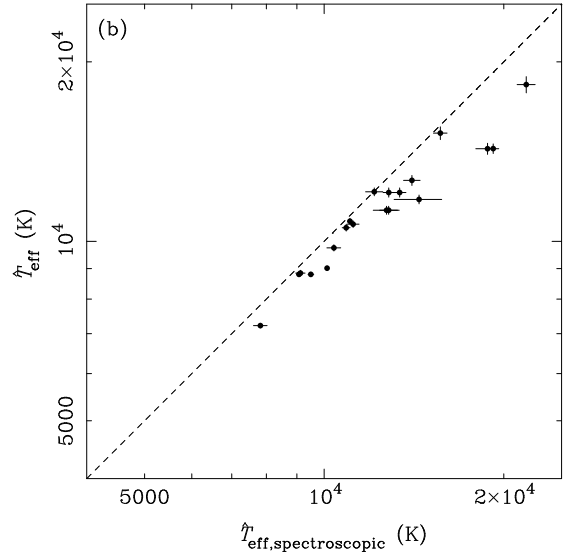}
\caption{The 
  estimated effective temperature, $\teffobs$, 
  of the \neisen\ of the \nhalo\ candidate halo WDs identified by \vidrih\
  for which there are spectroscopic fits in Eisenstein \etal\ (2006).
  Here the values estimated from spectra are compared to the 
  Bayesian estimates obtained from the 
  H atmosphere models using the 
  informative Galactic model prior.}
\label{figure:halospec_teff}
\end{figure}

Overall, the implication of these results is that many of the 
candidate halo stars do not actually have the high tangential velocities
that the data seem to imply.  
This hypothesis could be tested by obtaining 
follow-up observations, although higher precision photometry 
(other than $\uprime$ band) 
would be of little use, as the ambiguities are dominated by the 
fundamental WD model degeneracies, not the quality of the existing photometry.
Distances based on parallaxes would obviously be definitive,
but impractical with current observational resources. 
That leaves spectroscopic follow-up, 
although in many cases this is not necessary 
as \neisen\ of the \nhalo\ halo candidates already have 
spectroscopic fits presented by \cite{Eisenstein_etal:2006}.
These spectroscopic fits 
(only $\logg$ and $\teff$, but not $\dist$) 
are given in 
\tabl{haloeisen},
and compared to the Bayesian estimates of 
$\teff$ in 
\fig{halospec_teff}.

\begin{table*}
\begin{minipage}{165mm}
\centering
\scriptsize
\caption{Model selection quantities for the
  \nhalo\ \vidrih\ candidate halo WDs, where columns 2 and 3 are the 
  \bestfit\ $\logg$ values estimated using the uninformative $\loggpar$ 
  prior. UKIDSS data have been used where available. The final column 
  states whether the model selection result is driven by the data, 
  the informative $\loggpar$ prior, or the 10:1 model prior.}
\label{table:halomodel}
\begin{tabular}{cccrrcccl}
\hline
{object} & $\log g_\hydrogen$ & $\log g_\helium$ &  $\chi^2_{\rm min, \hydrogen}$ & $\chi^2_{\rm min, \helium}$ & $E_\hydrogen$ & $E_\helium$ & $P_\hydrogen$ & notes \\
\hline
SDSS~J000244.03$+$010945.8 & 7.57 & 7.83 &  16.5 &  14.1 & $1.2\times 10^{-02}$ & $1.2\times 10^{-01}$ & $0.47$ & data and $\loggpar$ prior \\
SDSS~J000557.24$+$001833.2 & 7.86 & 7.89 &   2.8 &   4.1 & $5.6\times 10^{-02}$ & $2.3\times 10^{-02}$ & $0.96$ & data and model prior \\
SDSS~J001306.23$+$005506.4 & 7.55 & 7.84 &  14.8 &  12.6 & $3.1\times 10^{-03}$ & $4.7\times 10^{-02}$ & $0.37$ & data and $\loggpar$ prior  \\
SDSS~J001518.33$+$010549.2 & 7.38 & 7.94 &  13.3 &   8.9 & $1.8\times 10^{-02}$ & $1.3$ & $0.11$ & data and $\loggpar$ prior \\
SDSS~J001838.54$+$005943.5 & 7.57 & 7.97 &  11.1 &  10.3 & $7.1\times 10^{-01}$ & $4.5$ & $0.58$ & model prior \\
SDSS~J002951.64$+$005623.6 & 7.74 & 7.87 &  10.5 &   8.6 & $7.1\times 10^{-01}$ & $4.4$ & $0.59$ & model prior \\
SDSS~J003054.06$+$001115.6 & 7.90 & 7.95 &  23.3 &  30.3 & $6.0\times 10^{-04}$ & $1.3\times 10^{-05}$ & $1.00$ & data, neither model good fit \\
SDSS~J003730.58$-$001657.8 & 7.81 & 7.88 &   0.9 &   3.4 & $1.9\times 10^{+01}$ & $4.5$ & $0.97$ & data and model prior \\
SDSS~J003813.53$-$000128.8 & 7.61 & 7.77 &   3.4 &   3.9 & $3.3\times 10^{-01}$ & $5.5\times 10^{-01}$ & $0.84$ & model prior \\
SDSS~J004214.88$+$001135.7 & 7.68 & 7.79 &  12.4 &  11.4 & $2.1\times 10^{-01}$ & $9.7\times 10^{-01}$ & $0.66$ & model prior \\
SDSS~J005906.78$+$001725.2 & 7.73 & 7.84 &   6.0 &   4.8 & $3.9\times 10^{-01}$ & $2.1$ & $0.62$ & model prior \\
SDSS~J010129.81$-$003041.7 & 7.54 & 7.86 &  16.0 &  15.2 & $3.2\times 10^{-02}$ & $1.8\times 10^{-01}$ & $0.62$ & model prior \\
SDSS~J010207.24$-$003259.7 & 7.93 & 7.79 &   0.5 &   1.1 & $1.1$ & $9.6\times 10^{-01}$ & $0.91$ & data, $\loggpar$ and model prior \\
SDSS~J010225.13$-$005458.4 & 7.65 & 7.96 &   0.9 &   0.2 & $3.9\times 10^{+01}$ & $1.2\times 10^{+02}$ & $0.75$ & model prior \\
SDSS~J014247.10$+$005228.4 & 7.46 & 7.87 &   2.1 &   0.4 & $6.6$ & $5.7\times 10^{+01}$ & $0.51$ & model prior \\
SDSS~J015227.57$-$002421.1 & 7.81 & 7.79 &   1.5 &   1.7 & $1.8\times 10^{+01}$ & $1.6\times 10^{+01}$ & $0.91$ & model prior \\
SDSS~J020241.81$-$005743.0 & 7.82 & 7.89 &   9.3 &  11.2 & $4.6\times 10^{-02}$ & $1.2\times 10^{-02}$ & $0.97$ & data and model prior \\
SDSS~J020729.85$+$000637.6 & 7.78 & 8.61 &   8.4 &  23.4 & $6.6\times 10^{-03}$ & $8.6\times 10^{-08}$ & $1.00$ & data \\
SDSS~J024529.69$-$004229.8 & 7.43 & 7.86 &   8.7 &   6.0 & $7.2\times 10^{-01}$ & $1.2\times 10^{+01}$ & $0.35$ & data and $\loggpar$ prior \\
SDSS~J024837.53$-$003123.9 & 8.08 & 8.10 &  15.3 &  31.6 & $4.2\times 10^{-05}$ & $2.6\times 10^{-08}$ & $1.00$ & data \\
SDSS~J025325.83$-$002751.5 & 7.13 & 7.93 &  97.3 & 118.3 & $4.1\times 10^{-24}$ & $6.2\times 10^{-26}$ & $1.00$ & data, neither model is a good fit \\
SDSS~J025531.00$-$005552.8 & 7.64 & 7.91 &   5.7 &   3.7 & $2.2$ & $1.6\times 10^{+01}$ & $0.55$ & model prior  \\
SDSS~J030433.61$-$002733.2 & 7.74 & 8.22 &  15.6 &  31.0 & $7.1\times 10^{-03}$ & $1.3\times 10^{-06}$ & $1.00$ & data and $\loggpar$ prior \\
SDSS~J211928.44$-$002632.9 & 7.41 & 8.06 &   7.7 &   4.7 & $7.8\times 10^{-01}$ & $1.3\times 10^{+01}$ & $0.36$ & data and $\loggpar$ prior \\
SDSS~J213641.39$+$010504.9 & 7.54 & 7.92 &  13.9 &  11.0 & $4.0\times 10^{-02}$ & $5.8\times 10^{-01}$ & $0.38$ & data and $\loggpar$ prior  \\
SDSS~J215138.09$+$003222.3 & 7.99 & 8.09 &   7.1 &  22.6 & $2.3\times 10^{-01}$ & $8.0\times 10^{-05}$ & $1.00$ & data, $\loggpar$ and model prior \\
SDSS~J223808.18$+$003247.9 & 7.64 & 8.43 &  40.2 &  75.4 & $1.7\times 10^{-09}$ & $9.2\times 10^{-18}$ & $1.00$ & data, neither model is a good fit \\
SDSS~J223815.97$-$011336.9 & 7.55 & 7.77 &  17.4 &  10.9 & $4.2\times 10^{-03}$ & $1.8\times 10^{-01}$ & $0.17$ &  data and $\loggpar$ prior \\
SDSS~J230534.79$-$010225.2 & 7.71 & 7.81 &   9.0 &   7.9 & $7.6\times 10^{-01}$ & $2.4$ & $0.74$ & model prior  \\
SDSS~J231626.98$+$004607.0 & 7.88 & 8.34 &   7.4 &   0.6 & $1.4\times 10^{-02}$ & $1.1\times 10^{-01}$ & $0.55$ & $\loggpar$ and model prior  \\
SDSS~J233227.63$-$010713.8 & 7.64 & 7.84 &   4.4 &   2.1 & $1.7$ & $1.1\times 10^{+01}$ & $0.57$ & model prior \\
SDSS~J233817.06$-$005720.1 & 7.94 & 7.85 &  17.5 &  21.4 & $7.5\times 10^{-03}$ & $1.9\times 10^{-03}$ & $0.97$ & data  \\
SDSS~J234110.13$+$003259.5 & 7.54 & 7.86 &   3.5 &   1.3 & $1.8$ & $2.4\times 10^{+01}$ & $0.40$ & data and $\loggpar$ prior \\
SDSS~J235138.85$+$002716.9 & 7.67 & 7.94 &  10.5 &  16.2 & $1.7\times 10^{-01}$ & $8.4\times 10^{-03}$ & $0.99$ & data and model prior \\
\hline
\end{tabular}
\end{minipage}
\end{table*}

The surface gravity estimates obtained with the non-informative 
prior reveal that the photometric data for the halo candidates 
are most consistent with low $\loggpar$,  
as expected given the combination of solid angle degeneracy
and high-$\vtanobs$ selection.  
There is, however, a suggestion that the true surface
gravities are systematically higher than the fiducial value of $\logg \simeq 7.9$,
a notion that is supported by the fact that the available spectroscopic 
$\logg$ fits are similarly discrepant.
The simplest explanation for this is that
the Gaussian $\logg$ prior given in \eq{loggprior} is overly prescriptive.
Correcting the tails of the 
prior would not alter the best fit values,
but would proportionally increase the tails of $\logg$ posteriors.
Given that the solid angle degeneracy ensures that a broad range of
$\logg$ values are consistent with the data,
the photometric measurements add minimal new information, 
and so the posterior peaks at the prior value for each WD in turn.
Some particular cases of this phenomenon are discussed in 
\sect{haloindividual}.

The spectroscopic and photometric
temperature estimates shown in \fig{halospec_teff} are 
reasonably consistent with each other, 
but the Bayesian values are systematically
lower than those from the spectroscopic fits.  
This is not the result of
the steep $\teff$ prior given in \eq{teffprior}, 
but is another, as yet unidentified, systematic difference between
the two approaches.  
The natural assumption would be that the spectroscopic fits are 
correct given the far greater available information; however,
as can be seen in \tabl{haloeisen}, many of these WD have
temperatures of 
$\teff \la 11,000 \unit{K}$
for which the spectroscopic models are biased
\citep{Eisenstein_etal:2006}.
A final possibility is, once again, a statistical explanation:
whereas there is no reason that the spectroscopic data
would be at the margins of the noise distribution, 
it is plausible that the photometric data on the same objects
are biased, as it was these data that were used to select the sample.


\subsubsection{Notes on individual objects}
\label{section:haloindividual}

\begin{figure*}
\includegraphics[width=\figurewidth]
  {\figdir 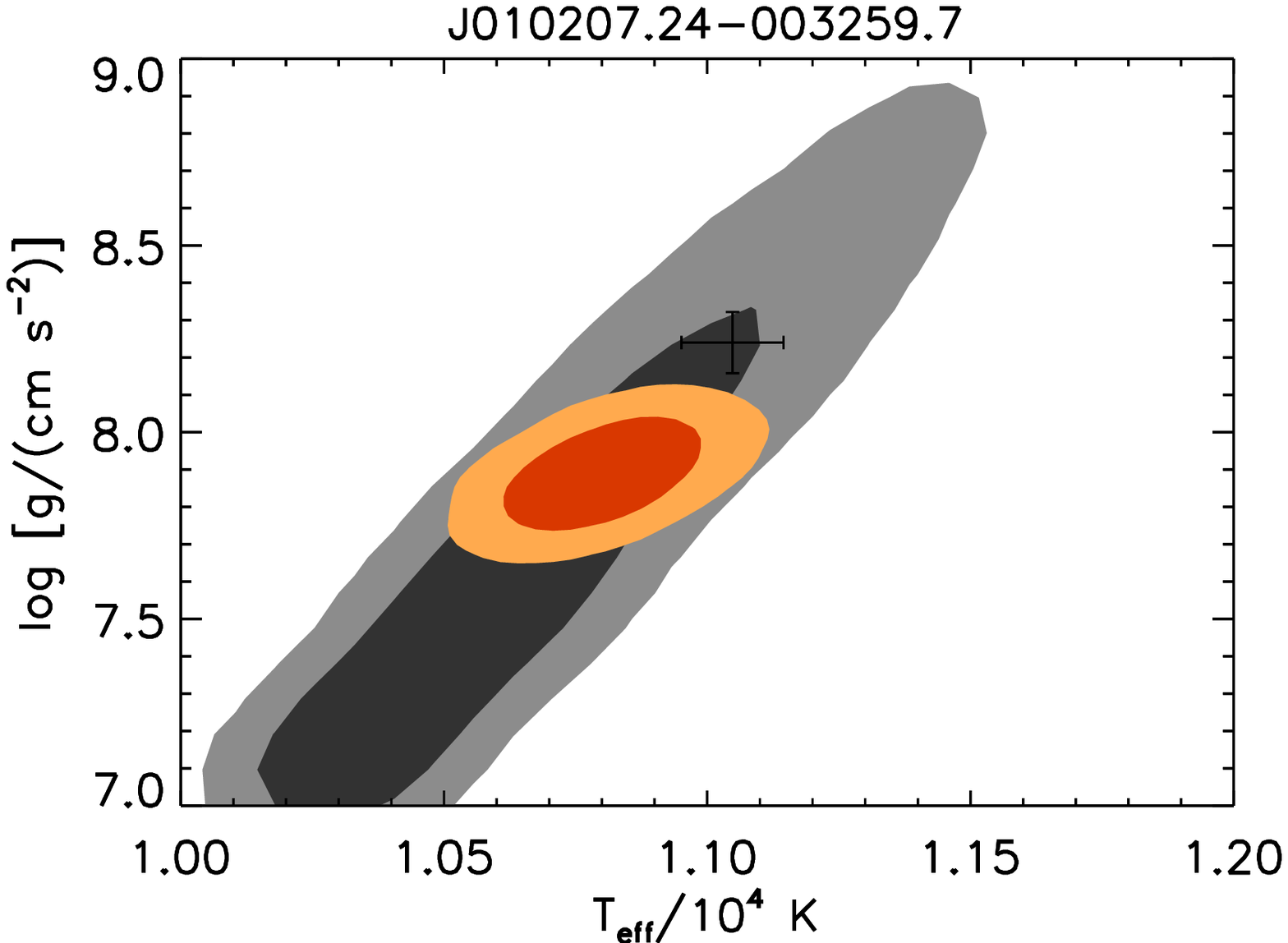}
\includegraphics[width=\figurewidth]
  {\figdir 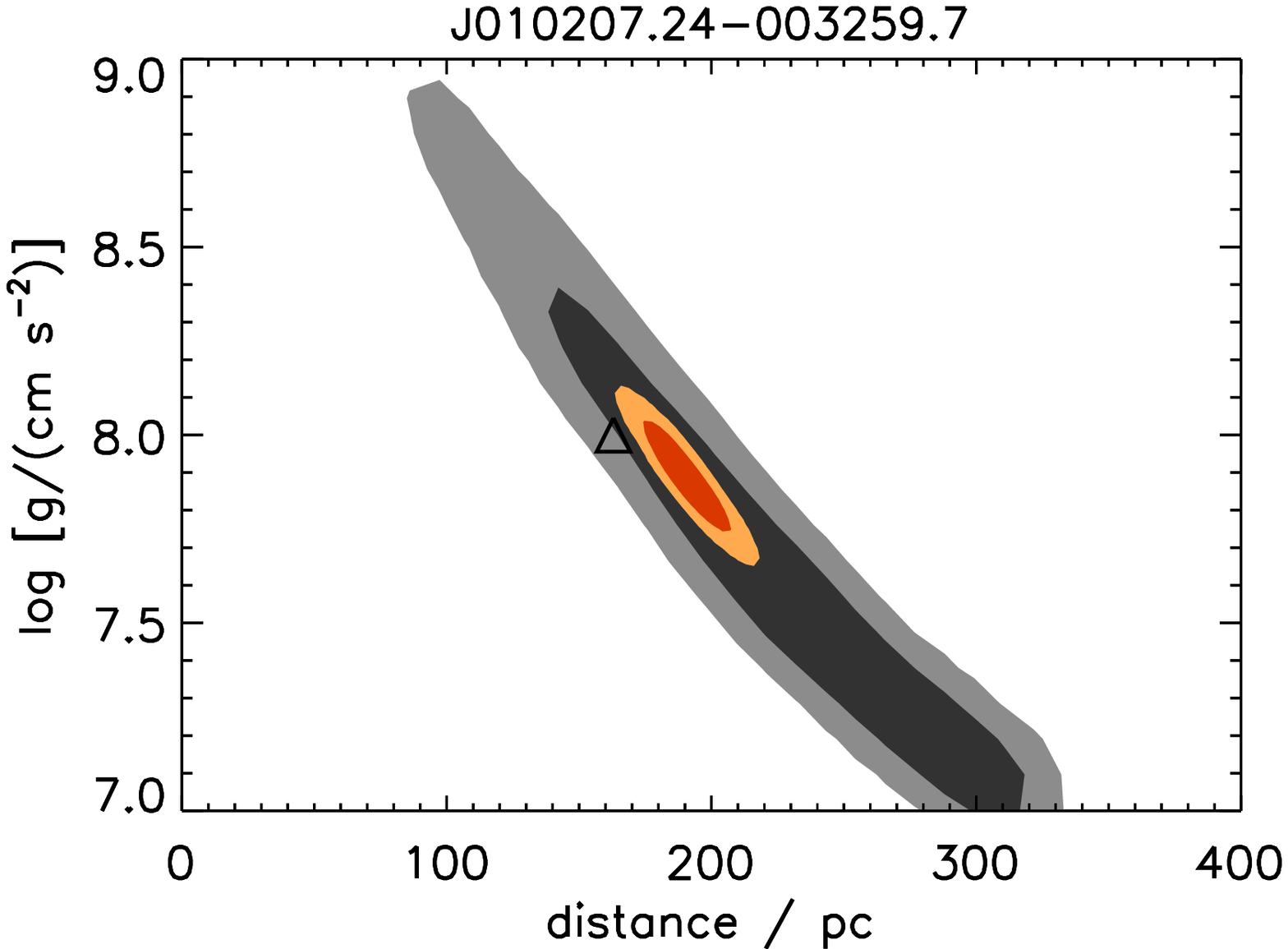}
\caption{Marginalised posterior distributions
  of the model parameters for the candidate halo WD
  SDSS~J010207.24$-$003259.7.
  The point with the error bar in the left panel is the
  Eisenstein \etal\ (2006) fit to the SDSS spectrum of this WD.
  For both the non-informative prior (grey) and the informative prior
  (orange) the coloured regions enclose 68 \percent\ and 95 \percent\
  of the posterior probability in these two parameters.
  The open triangle in the right panel is the Harris \etal\ (2006) fit
  assuming $\logg = 8.0$.}
\label{figure:halo13}
\end{figure*}

The results for most of the \nhalo\ candidate halo WDs are 
explained by the general arguments given above, but it is 
useful to investigate a few of sources in greater detail.

Of particular interest are the three objects for which spectroscopic 
surface gravity estimates given by \cite{Eisenstein_etal:2006} 
are significantly higher than the fiducial value of $\logg \simeq 7.9$: 
SDSS~J005906.78$+$001725.2, with $\loggobs = 8.53 \pm 0.38$;
SDSS~J010225.13$-$005458.4, with $\loggobs = 8.95 \pm 0.06$; and 
SDSS~J014247.10$+$005228.4, with $\loggobs = 8.82 \pm 0.11$.
The first of these was identified as a DZ type WD by \cite{Kleinman:2004}, 
and it is perhaps not surprising that the DA and DB atmosphere fits 
presented here and in \vidrih\ do not match the 
spectroscopic estimate. 
The latter two objects were identified as DA type WDs by 
\cite{Eisenstein_etal:2006}. 
There are no UKIDSS measurements for these two stars,
and the goodness--of--fit to the SDSS photometry shows nothing unusual. 
The implication is 
that these objects have intrinsically unusual surface gravity, and 
that the photometry is insufficient to add any information to the 
expectations of the $\logg$ prior. 
It further follows that, because of the direction of the degeneracy 
between $\loggpar$ and $\dist$, a higher value of $\logg$ would imply a 
significantly lower distance, and hence, lower $\vtan$. 

\begin{figure}
\includegraphics[width=\figurewidth]  
  {\figdir 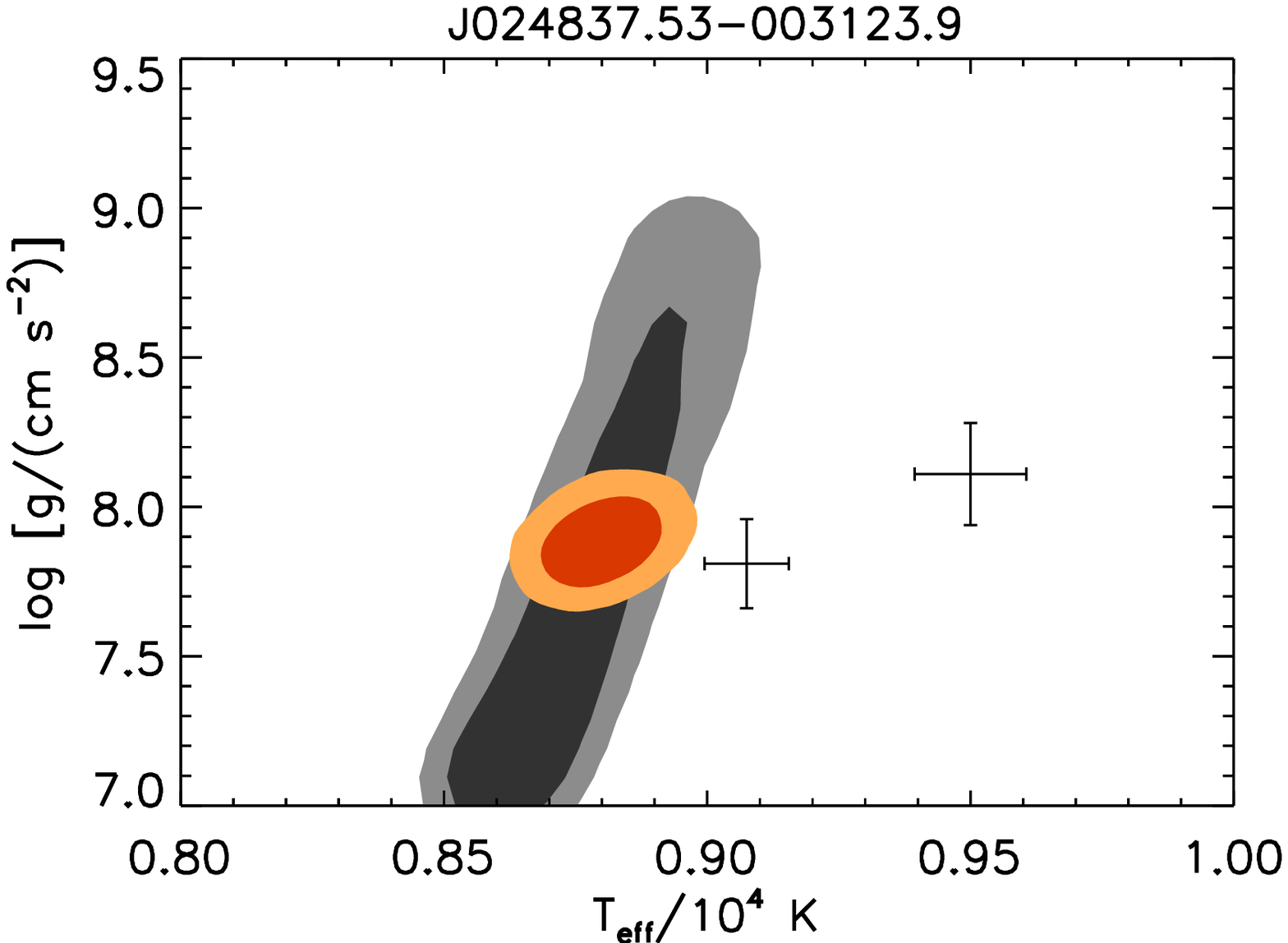}
\caption{Same as \fig{halo13}(a), but for the candidate halo WD 
  SDSS~J024837.53$-$003123.9.
  The two points with errors are the 
  Eisenstein \etal\ (2006) fits to the two SDSS spectra of this WD.}
\label{figure:halo20}
\end{figure}

There are also three WDs seen in \fig{halovtan}~(b) with strikingly 
high Bayesian estimates for $\vtan$ compared with the \vidrih\ estimates:  
SDSS~J010207.24$-$003259.7, with $\vtanobs \simeq 334 \kms$, 
SDSS~J024837.53$-$003123.9, with $\vtanobs \simeq 328 \kms$, 
and SDSS~J223808.18$+$003247.9, with $\vtanobs \simeq 289 \kms$. 
There are UKIDSS observations for all three stars and, fortuitously, 
spectroscopic $\logg$ estimates given by \cite{Eisenstein_etal:2006} 
for the first two, 
the marginalised posterior distributions for which are shown in 
\figs{halo13} and \ref{figure:halo20}, respectively.
For SDSS~J010207.24$-$003259.7 the 
spectroscopic estimate of $\loggobs = 8.24 \pm 0.08$ implies
that it is probably not a viable halo candidate.
The spectroscopic $\logg$ estimates 
for SDSS~J024837.53$-$003123.9
are consistent with the Bayesian 
estimates, and hence imply 
this WD may have genuinely unusual dynamics. 
The fit for the third object, SDSS~J223808.18$+$003247.9,
which does not have spectroscopic data,
is disastrously bad for both H and He models, with or
without UKIDSS data. 

SDSS~J024837.53$-$003123.9 has duplicate observations in the 
\cite{Eisenstein_etal:2006} sample (see the left panel of \fig{halo20}), 
and it is interesting to note that the error bars given for $\teff$ 
for these repeated observations are marginally inconsistent with each other 
at face value (as well as being systematically higher than the Bayesian 
estimate, as mentioned above). 
The temperature uncertainties from the photometry are 
comparably precise to those from spectroscopy, 
although the surface gravity is only constrained by the 
spectroscopic data.

Although the model constraints on SDSS~J030433.61$-$002733.2 
are not interesting in themselves it is important to note 
that the UKIDSS data associated with this object in \vidrih\
is results from an
incorrect cross-match to a much cooler star $3\farcs5$ away.



\subsection{Candidate ultra-cool white dwarfs}
\label{section:ultracool}

\begin{table*}
\begin{minipage}{150mm}
\centering
\scriptsize
\caption{Estimated H atmosphere model parameters of the 
  \nultracool\ \vidrih\ candidate ultra-cool WDs,
  obtained using the informative prior defined in \eq{prior}.}
\label{table:ultracoolh}
\begin{tabular}{ccccccc}
\hline
object & $\teff$ & $\logg$ & $M$       & $T$  & $\dist$ & $\vtan$ \\
       & (K)     &         & $(\msun)$ & (yr) & (pc) & (km s$^{-1}$) \\
\hline
SDSS J001107.57$-$003102.8 & $( 3.64 \pm 0.04) \times 10^{ 3} $ & $ 7.36 \pm 0.06 $ & $ 0.27 \pm 0.02 $ & $( 4.28 \pm 0.37) \times 10^{ 9} $ & $   105 \pm  3 $ & $    28 \pm  3 $ \\
SDSS J004843.28$-$003820.0 & $( 3.72 \pm 0.05) \times 10^{ 3} $ & $ 7.53 \pm 0.09 $ & $ 0.34 \pm 0.04 $ & $( 5.07 \pm 0.57) \times 10^{ 9} $ & $   146 \pm  6 $ & $    37 \pm  6 $ \\
SDSS J012102.99$-$003833.6 & $( 3.93 \pm 0.04) \times 10^{ 3} $ & $ 7.75 \pm 0.07 $ & $ 0.45 \pm 0.03 $ & $( 6.24 \pm 0.66) \times 10^{ 9} $ & $    67 \pm  3 $ & $    39 \pm  2 $ \\
SDSS J013302.17$+$010201.3 & $( 3.62 \pm 0.10) \times 10^{ 3} $ & $ 7.62 \pm 0.10 $ & $ 0.38 \pm 0.05 $ & $( 6.13 \pm 0.75) \times 10^{ 9} $ & $   145 \pm  7 $ & $    34 \pm  6 $ \\
SDSS J030144.09$-$004439.5 & $( 3.62 \pm 0.05) \times 10^{ 3} $ & $ 7.77 \pm 0.05 $ & $ 0.46 \pm 0.03 $ & $( 7.28 \pm 0.52) \times 10^{ 9} $ & $    50 \pm  2 $ & $   132 \pm  4 $ \\
SDSS J204332.97$+$011436.2 & $( 3.56 \pm 0.07) \times 10^{ 3} $ & $ 7.62 \pm 0.09 $ & $ 0.38 \pm 0.05 $ & $( 6.35 \pm 0.73) \times 10^{ 9} $ & $   139 \pm  7 $ & $    38 \pm  9 $ \\
SDSS J205010.17$+$003233.7 & $( 3.85 \pm 0.14) \times 10^{ 3} $ & $ 7.85 \pm 0.10 $ & $ 0.50 \pm 0.05 $ & $( 7.55 \pm 1.14) \times 10^{ 9} $ & $    92 \pm  7 $ & $    24 \pm  6 $ \\
SDSS J205132.05$+$000353.6 & $( 3.46 \pm 0.04) \times 10^{ 3} $ & $ 7.05 \pm 0.04 $ & $ 0.17 \pm 0.01 $ & $( 3.18 \pm 0.17) \times 10^{ 9} $ & $    69 \pm  2 $ & $   331 \pm  9 $ \\
SDSS J210742.26$-$002354.1 & $( 3.49 \pm 0.08) \times 10^{ 3} $ & $ 7.77 \pm 0.09 $ & $ 0.46 \pm 0.05 $ & $( 7.86 \pm 0.81) \times 10^{ 9} $ & $   109 \pm  5 $ & $    42 \pm  6 $ \\
SDSS J212216.01$-$010715.2 & $( 4.09 \pm 0.06) \times 10^{ 3} $ & $ 7.91 \pm 0.10 $ & $ 0.53 \pm 0.05 $ & $( 7.53 \pm 1.23) \times 10^{ 9} $ & $   105 \pm  7 $ & $    84 \pm  7 $ \\
SDSS J212930.25$-$003411.5 & $( 3.93 \pm 0.06) \times 10^{ 3} $ & $ 7.98 \pm 0.10 $ & $ 0.57 \pm 0.06 $ & $( 8.56 \pm 0.98) \times 10^{ 9} $ & $    76 \pm  6 $ & $   154 \pm 11 $ \\
SDSS J214108.42$+$002629.5 & $( 3.60 \pm 0.10) \times 10^{ 3} $ & $ 7.60 \pm 0.10 $ & $ 0.37 \pm 0.05 $ & $( 6.03 \pm 0.73) \times 10^{ 9} $ & $   150 \pm  8 $ & $    38 \pm  8 $ \\
SDSS J220455.03$-$001750.6 & $( 3.76 \pm 0.14) \times 10^{ 3} $ & $ 7.97 \pm 0.10 $ & $ 0.57 \pm 0.06 $ & $( 8.97 \pm 1.06) \times 10^{ 9} $ & $    81 \pm  7 $ & $    41 \pm  4 $ \\
SDSS J223105.29$+$004941.9 & $( 3.72 \pm 0.04) \times 10^{ 3} $ & $ 7.49 \pm 0.07 $ & $ 0.32 \pm 0.03 $ & $( 4.78 \pm 0.41) \times 10^{ 9} $ & $   113 \pm  3 $ & $    61 \pm  3 $ \\
SDSS J223520.19$-$003623.6 & $( 3.74 \pm 0.10) \times 10^{ 3} $ & $ 7.91 \pm 0.09 $ & $ 0.53 \pm 0.05 $ & $( 8.41 \pm 1.17) \times 10^{ 9} $ & $   110 \pm  7 $ & $    73 \pm  6 $ \\
SDSS J223715.32$-$002939.2 & $( 3.30 \pm 0.06) \times 10^{ 3} $ & $ 7.72 \pm 0.08 $ & $ 0.43 \pm 0.04 $ & $( 8.05 \pm 0.60) \times 10^{ 9} $ & $    99 \pm  4 $ & $    30 \pm  3 $ \\
SDSS J223954.07$+$001849.2 & $( 4.22 \pm 0.06) \times 10^{ 3} $ & $ 7.79 \pm 0.09 $ & $ 0.47 \pm 0.05 $ & $( 6.00 \pm 0.91) \times 10^{ 9} $ & $    79 \pm  4 $ & $    41 \pm  3 $ \\
SDSS J224206.19$+$004822.7 & $( 3.40 \pm 0.03) \times 10^{ 3} $ & $ 8.29 \pm 0.06 $ & $ 0.77 \pm 0.04 $ & $( 1.07 \pm 0.01) \times 10^{10} $ & $    24 \pm  1 $ & $    18 \pm  1 $ \\
SDSS J224845.93$-$005407.0 & $( 4.02 \pm 0.05) \times 10^{ 3} $ & $ 7.93 \pm 0.10 $ & $ 0.54 \pm 0.06 $ & $( 7.90 \pm 1.22) \times 10^{ 9} $ & $    91 \pm  6 $ & $    88 \pm  7 $ \\
SDSS J225244.51$+$000918.6 & $( 3.73 \pm 0.14) \times 10^{ 3} $ & $ 7.94 \pm 0.10 $ & $ 0.55 \pm 0.06 $ & $( 8.70 \pm 1.25) \times 10^{ 9} $ & $    96 \pm  9 $ & $    39 \pm  5 $ \\
SDSS J232115.67$+$010223.8 & $( 4.46 \pm 0.03) \times 10^{ 3} $ & $ 7.94 \pm 0.10 $ & $ 0.55 \pm 0.06 $ & $( 6.79 \pm 1.07) \times 10^{ 9} $ & $    53 \pm  3 $ & $    68 \pm  5 $ \\
SDSS J233055.19$+$002852.2 & $( 4.23 \pm 0.03) \times 10^{ 3} $ & $ 7.89 \pm 0.08 $ & $ 0.52 \pm 0.04 $ & $( 7.08 \pm 0.96) \times 10^{ 9} $ & $    49 \pm  2 $ & $    39 \pm  2 $ \\
SDSS J233818.56$-$004146.2 & $( 3.70 \pm 0.06) \times 10^{ 3} $ & $ 7.72 \pm 0.09 $ & $ 0.43 \pm 0.04 $ & $( 6.66 \pm 0.82) \times 10^{ 9} $ & $   122 \pm  6 $ & $    82 \pm  6 $ \\
SDSS J234646.06$-$003527.6 & $( 3.70 \pm 0.06) \times 10^{ 3} $ & $ 7.55 \pm 0.09 $ & $ 0.35 \pm 0.04 $ & $( 5.24 \pm 0.60) \times 10^{ 9} $ & $   143 \pm  6 $ & $    30 \pm  5 $ \\
\hline
\end{tabular}
\end{minipage}
\end{table*}

\begin{table*}
\begin{minipage}{151mm}
\centering
\scriptsize
\caption{Estimated He atmosphere model parameters of the   
  \nultracool\ \vidrih\ candidate ultra-cool WDs,
  obtained using the informative prior defined in \eq{prior}.}
\label{table:ultracoolhe}
\begin{tabular}{ccccccc}
\hline
object & $\teff$ & $\logg$ & $M$       & $T$  & $\dist$ & $\vtan$ \\
       & (K)     &         & $(\msun)$ & (yr) & (pc) & (km s$^{-1}$) \\
\hline
SDSS J001107.57$-$003102.8 & $( 4.49 \pm 0.04) \times 10^{ 3} $ & $ 7.99 \pm 0.11 $ & $ 0.57 \pm 0.06 $ & $( 6.68 \pm 0.78) \times 10^{ 9} $ & $   106 \pm  9 $ & $    29 \pm  4 $ \\
SDSS J004843.28$-$003820.0 & $( 4.42 \pm 0.04) \times 10^{ 3} $ & $ 7.93 \pm 0.10 $ & $ 0.53 \pm 0.05 $ & $( 6.41 \pm 1.00) \times 10^{ 9} $ & $   154 \pm 11 $ & $    39 \pm  7 $ \\
SDSS J012102.99$-$003833.6 & $( 4.65 \pm 0.02) \times 10^{ 3} $ & $ 7.92 \pm 0.10 $ & $ 0.53 \pm 0.05 $ & $( 5.83 \pm 1.00) \times 10^{ 9} $ & $    83 \pm  5 $ & $    48 \pm  4 $ \\
SDSS J013302.17$+$010201.3 & $( 4.35 \pm 0.05) \times 10^{ 3} $ & $ 7.91 \pm 0.10 $ & $ 0.53 \pm 0.05 $ & $( 6.42 \pm 1.07) \times 10^{ 9} $ & $   159 \pm 12 $ & $    38 \pm  7 $ \\
SDSS J030144.09$-$004439.5 & $( 4.60 \pm 0.02) \times 10^{ 3} $ & $ 7.90 \pm 0.10 $ & $ 0.52 \pm 0.05 $ & $( 5.74 \pm 1.05) \times 10^{ 9} $ & $    70 \pm  4 $ & $   185 \pm 12 $ \\
SDSS J204332.97$+$011436.2 & $( 4.33 \pm 0.05) \times 10^{ 3} $ & $ 7.90 \pm 0.10 $ & $ 0.52 \pm 0.05 $ & $( 6.39 \pm 1.09) \times 10^{ 9} $ & $   155 \pm 12 $ & $    43 \pm 10 $ \\
SDSS J205010.17$+$003233.7 & $( 4.60 \pm 0.06) \times 10^{ 3} $ & $ 7.90 \pm 0.10 $ & $ 0.52 \pm 0.05 $ & $( 5.75 \pm 1.05) \times 10^{ 9} $ & $   120 \pm  9 $ & $    31 \pm  7 $ \\
SDSS J205132.05$+$000353.6 & $( 4.15 \pm 0.04) \times 10^{ 3} $ & $ 7.92 \pm 0.10 $ & $ 0.53 \pm 0.05 $ & $( 6.88 \pm 1.10) \times 10^{ 9} $ & $    54 \pm  4 $ & $   262 \pm 21 $ \\
SDSS J210742.26$-$002354.1 & $( 4.52 \pm 0.03) \times 10^{ 3} $ & $ 7.89 \pm 0.10 $ & $ 0.51 \pm 0.05 $ & $( 5.85 \pm 1.06) \times 10^{ 9} $ & $   154 \pm 11 $ & $    60 \pm  8 $ \\
SDSS J212216.01$-$010715.2 & $( 4.73 \pm 0.03) \times 10^{ 3} $ & $ 7.89 \pm 0.10 $ & $ 0.51 \pm 0.05 $ & $( 5.44 \pm 1.00) \times 10^{ 9} $ & $   142 \pm  9 $ & $   115 \pm  9 $ \\
SDSS J212930.25$-$003411.5 & $( 4.71 \pm 0.03) \times 10^{ 3} $ & $ 7.91 \pm 0.10 $ & $ 0.53 \pm 0.05 $ & $( 5.66 \pm 1.01) \times 10^{ 9} $ & $   110 \pm  7 $ & $   224 \pm 15 $ \\
SDSS J214108.42$+$002629.5 & $( 4.32 \pm 0.05) \times 10^{ 3} $ & $ 7.91 \pm 0.10 $ & $ 0.53 \pm 0.05 $ & $( 6.46 \pm 1.09) \times 10^{ 9} $ & $   162 \pm 12 $ & $    41 \pm  9 $ \\
SDSS J220455.03$-$001750.6 & $( 4.66 \pm 0.03) \times 10^{ 3} $ & $ 7.89 \pm 0.10 $ & $ 0.51 \pm 0.05 $ & $( 5.54 \pm 1.01) \times 10^{ 9} $ & $   124 \pm  8 $ & $    63 \pm  5 $ \\
SDSS J223105.29$+$004941.9 & $( 4.55 \pm 0.03) \times 10^{ 3} $ & $ 7.96 \pm 0.10 $ & $ 0.55 \pm 0.06 $ & $( 6.35 \pm 0.89) \times 10^{ 9} $ & $   123 \pm  9 $ & $    67 \pm  6 $ \\
SDSS J223520.19$-$003623.6 & $( 4.64 \pm 0.03) \times 10^{ 3} $ & $ 7.90 \pm 0.10 $ & $ 0.52 \pm 0.05 $ & $( 5.69 \pm 1.05) \times 10^{ 9} $ & $   160 \pm 11 $ & $   106 \pm  8 $ \\
SDSS J223715.32$-$002939.2 & $( 4.54 \pm 0.03) \times 10^{ 3} $ & $ 7.86 \pm 0.10 $ & $ 0.50 \pm 0.05 $ & $( 5.61 \pm 1.00) \times 10^{ 9} $ & $   149 \pm 10 $ & $    46 \pm  6 $ \\
SDSS J223954.07$+$001849.2 & $( 4.60 \pm 0.04) \times 10^{ 3} $ & $ 7.90 \pm 0.09 $ & $ 0.52 \pm 0.05 $ & $( 5.75 \pm 1.02) \times 10^{ 9} $ & $    89 \pm  6 $ & $    46 \pm  3 $ \\
SDSS J224206.19$+$004822.7 & $( 4.84 \pm 0.02) \times 10^{ 3} $ & $ 7.66 \pm 0.09 $ & $ 0.40 \pm 0.05 $ & $( 3.48 \pm 0.63) \times 10^{ 9} $ & $    65 \pm  4 $ & $    50 \pm  3 $ \\
SDSS J224845.93$-$005407.0 & $( 4.72 \pm 0.03) \times 10^{ 3} $ & $ 7.93 \pm 0.10 $ & $ 0.53 \pm 0.05 $ & $( 5.77 \pm 0.96) \times 10^{ 9} $ & $   125 \pm  9 $ & $   121 \pm  9 $ \\
SDSS J225244.51$+$000918.6 & $( 4.65 \pm 0.03) \times 10^{ 3} $ & $ 7.89 \pm 0.10 $ & $ 0.52 \pm 0.05 $ & $( 5.60 \pm 1.04) \times 10^{ 9} $ & $   145 \pm 10 $ & $    59 \pm  7 $ \\
SDSS J232115.67$+$010223.8 & $( 4.87 \pm 0.02) \times 10^{ 3} $ & $ 7.92 \pm 0.09 $ & $ 0.53 \pm 0.05 $ & $( 5.40 \pm 0.92) \times 10^{ 9} $ & $    65 \pm  4 $ & $    84 \pm  5 $ \\
SDSS J233055.19$+$002852.2 & $( 4.76 \pm 0.02) \times 10^{ 3} $ & $ 7.91 \pm 0.10 $ & $ 0.52 \pm 0.05 $ & $( 5.54 \pm 1.01) \times 10^{ 9} $ & $    62 \pm  4 $ & $    49 \pm  3 $ \\
SDSS J233818.56$-$004146.2 & $( 4.55 \pm 0.03) \times 10^{ 3} $ & $ 7.93 \pm 0.10 $ & $ 0.54 \pm 0.05 $ & $( 6.14 \pm 1.01) \times 10^{ 9} $ & $   153 \pm 11 $ & $   103 \pm  9 $ \\
SDSS J234646.06$-$003527.6 & $( 4.42 \pm 0.04) \times 10^{ 3} $ & $ 7.93 \pm 0.10 $ & $ 0.54 \pm 0.05 $ & $( 6.42 \pm 0.99) \times 10^{ 9} $ & $   154 \pm 11 $ & $    32 \pm  6 $ \\
\hline
\end{tabular}
\end{minipage}
\end{table*}

\begin{figure}
\includegraphics[width=\figurewidth]{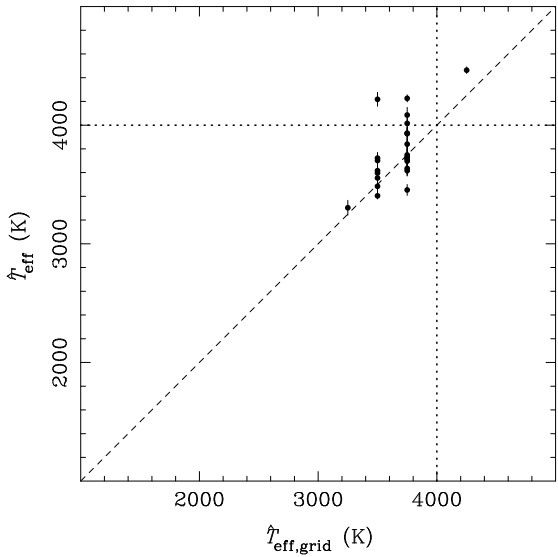}
\caption{The estimated effective temperature, $\teffobs$, of the
  \nultracool\ candidate ultra-cool WDs as found by \vidrih\ 
  and obtained using the Bayesian method described here 
  (for H atmosphere models,
  with the Galactic model priors).
  The selection criterion as an ultra-cool WD, $\teffobs \leq 4000\unit{K}$,
  adopted by \vidrih\ is shown as the vertical and horizontal
  dotted lines.  One source, SDSS~J232115.67$+$010223.8, is listed in this 
  sample despite an estimated effective temperature above this nominal limit.}
\label{figure:ultracoolteff}
\end{figure}

The \nultracool\ candidate ultra-cool WDs identified by 
\vidrih\ were analysed using the techniques described in 
\sect{parameterest} and \sect{modelselection}, giving parameter estimates 
and evidence values for both H and He atmosphere models.
The results obtained using the informative prior defined in \sect{wdpop}
are presented in \tabls{ultracoolh} and \ref{table:ultracoolhe}.
The Bayesian parameter fits are compared to those obtained by \vidrih\ 
in \fig{ultracoolteff} for the informative prior.

For the ultra-cool WDs, the temperature comparison shown in 
\fig{ultracoolteff} is the critical test as it was on the basis of 
$\teffobsgrid$ that these \nultracool\ objects were selected.
The discretisation on the horizontal axis is again due to the 
grid-based fitting of \vidrih.
The slight tendency for the Bayesian fits to have a higher 
$\teffobs$ is due to the selection effect that resulted in
the more pronounced distance bias discussed in \sect{halo}.
  
The various distance estimates for the ultra-cool WDs
are also in broad agreement.
Comparing the results obtained using the non-informative prior 
with those provided by \vidrih\ reveals only a small
scatter due to the \bestfit\ models lying between the model grid points.
The Bayesian distance estimates obtained from the informative prior 
and those given by \vidrih\ also exhibit systematic differences,
with the former tending towards the average
value of $\distobs \simeq 100 \unit{pc}$.
  

Given that the Bayesian distance estimates are in agreement with those of 
\vidrih, it is not surprising that the same holds for
the tangential velocity estimates.
The only significant exception to this is SDSS~J205132.05$+$000353.6,
for which the H fit shown is probably inappropriate, confirming the 
findings of \vidrih\ that this WD has a He atmosphere. 
This object is discussed further in \sect{ultracoolindividual}.
  


\begin{table*}
\begin{minipage}{170mm}
\centering
\scriptsize
\caption{Model selection quantities for the
  \nultracool\ \vidrih\ candidate ultra-cool WDs, where columns 2 and 
  3 are the best fit $\logg$ values estimated using the uninformative 
  $\loggpar$ prior. UKIDSS data have been used where available. The final 
  column states whether the model selection probability of $\hydrogen$ 
  versus $\helium$ is driven by the data, the informative $\loggpar$ prior, 
  or the 10:1 model prior.}
\label{table:ultracoolmodel}
\begin{tabular}{cccrrcccl}
\hline
{object} & $\log g_\hydrogen$ & $\log g_\helium$ &  $\chi^2_{\rm min, \hydrogen}$ & $\chi^2_{\rm min, \helium}$ & $E_\hydrogen$ & $E_\helium$ & $P_\hydrogen$ & notes \\
\hline
SDSS~J001107.57$-$003102.8 & 7.05 & 8.91 &  16.3 &  12.0 & $2.0\times 10^{-16}$ & $4.4\times 10^{-10}$ & $4.2\times 10^{-06}$ & data \\
SDSS~J004843.28$-$003820.0 & 7.05 & 8.80 &  44.3 &  22.7 & $3.6\times 10^{-18}$ & $2.7\times 10^{-07}$ & $1.2\times 10^{-10}$ & data, neither model good fit \\
SDSS~J012102.99$-$003833.6 & 7.59 & 8.21 &  29.8 &   4.7 & $4.1\times 10^{-11}$ & $2.6\times 10^{-05}$ & $1.4\times 10^{-05}$ & data \\
SDSS~J013302.17$+$010201.3 & 7.07 & 8.62 &  34.6 &  21.6 & $5.2\times 10^{-13}$ & $4.9\times 10^{-05}$ & $9.5\times 10^{-08}$ & data, neither model good fit \\
SDSS~J030144.09$-$004439.5 & 7.72 & 7.90 &  12.7 &  16.3 & $2.6\times 10^{-06}$ & $1.2\times 10^{-06}$ & $0.95$ & data and model prior \\
SDSS~J204332.97$+$011436.2 & 7.07 & 8.46 &  31.6 &  15.8 & $2.2\times 10^{-12}$ & $3.3\times 10^{-03}$ & $6.1\times 10^{-09}$ & data and $\loggpar$ prior \\
SDSS~J205010.17$+$003233.7 & 7.38 & 8.20 &   3.8 &   7.0 & $1.8\times 10^{+02}$ & $2.4\times 10^{+01}$ & $0.99$ & data and model prior \\
SDSS~J205132.05$+$000353.6 & 7.01 & 8.85 & 272.6 &  22.5 & $0.0$ & $8.8\times 10^{-09}$ & $0.00$ & data, neither model good fit \\
SDSS~J210742.26$-$002354.1 & 7.45 & 7.75 &   7.1 &   0.9 & $1.9\times 10^{-02}$ & $5.4$ & $3.1\times 10^{-02}$ & data and $\loggpar$ prior  \\
SDSS~J212216.01$-$010715.2 & 8.46 & 7.91 &   9.2 &   1.6 & $2.5\times 10^{-03}$ & $1.3$ & $1.8\times 10^{-02}$ & data and $\loggpar$ prior \\
SDSS~J212930.25$-$003411.5 & 8.31 & 8.45 &   5.7 &   0.2 & $7.8\times 10^{-04}$ & $4.0\times 10^{-02}$ & $0.15$ &  data \\
SDSS~J214108.42$+$002629.5 & 7.06 & 8.68 &  35.5 &  20.1 & $8.8\times 10^{-14}$ & $7.9\times 10^{-05}$ & $1.0\times 10^{-08}$ & data, neither model good fit \\
SDSS~J220455.03$-$001750.6 & 8.15 & 7.83 &  23.5 &   7.4 & $1.3\times 10^{-06}$ & $2.5\times 10^{-02}$ & $4.7\times 10^{-04}$ & data \\
SDSS~J223105.29$+$004941.9 & 7.07 & 8.84 &  10.2 &   4.0 & $2.3\times 10^{-12}$ & $1.7\times 10^{-06}$ & $1.2\times 10^{-05}$ & data \\
SDSS~J223520.19$-$003623.6 & 7.87 & 8.15 &   7.5 &   1.7 & $8.9\times 10^{-02}$ & $6.8\times 10^{-01}$ & $0.54$ & model prior and $\loggpar$ prior  \\
SDSS~J223715.32$-$002939.2 & 7.23 & 7.44 &  19.6 &  15.3 & $5.2\times 10^{-07}$ & $7.0\times 10^{-04}$ & $6.7\times 10^{-03}$ & data and $\loggpar$ prior \\
SDSS~J223954.07$+$001849.2 & 7.28 & 8.01 &  34.8 &  19.6 & $1.3\times 10^{-10}$ & $5.3\times 10^{-06}$ & $2.1\times 10^{-04}$ &  data and $\loggpar$ prior \\
SDSS~J224206.19$+$004822.7 & 8.58 & 7.07 &  44.8 & 286.1 & $5.0\times 10^{-21}$ & $0.0$ & $1.00$ &  data, neither model good fit \\
SDSS~J224845.93$-$005407.0 & 8.09 & 8.48 &  14.7 &   8.9 & $4.4\times 10^{-05}$ & $4.4\times 10^{-05}$ & $0.90$ & model prior and $\loggpar$ prior  \\
SDSS~J225244.51$+$000918.6 & 8.07 & 7.96 &   8.3 &   1.9 & $2.9\times 10^{-02}$ & $1.3$ & $0.17$ &  data and $\loggpar$ prior \\
SDSS~J232115.67$+$010223.8 & 8.46 & 8.43 &  19.1 &   4.8 & $6.2\times 10^{-08}$ & $9.2\times 10^{-05}$ & $6.1\times 10^{-03}$ & data \\
SDSS~J233055.19$+$002852.2 & 7.89 & 8.08 &  30.4 &   6.0 & $1.7\times 10^{-10}$ & $1.6\times 10^{-05}$ & $9.3\times 10^{-05}$ &  data \\
SDSS~J233818.56$-$004146.2 & 7.21 & 8.76 &  11.2 &  13.3 & $3.0\times 10^{-05}$ & $1.5\times 10^{-05}$ & $0.95$ & model prior and $\loggpar$ prior \\
SDSS~J234646.06$-$003527.6 & 7.06 & 8.82 &  30.7 &  22.8 & $1.4\times 10^{-14}$ & $9.8\times 10^{-08}$ & $1.3\times 10^{-06}$ & data, neither model good fit \\
\hline
\end{tabular}
\end{minipage}
\end{table*}

The model comparison results for the ultra-cool sample are
summarised in \tabl{ultracoolmodel}.
The most striking result is that 
18 out of the \nultracool\ candidate ultra-cool WDs have a high probability of 
being He stars, despite the strong model prior that H
stars should be $\sim 10$ times more common.
The same conclusion is also implied by the optical data alone,
although it is perhaps more revealing that several stars are poor
fits to both the H and He models.
It is also interesting to note that the \bestfit\ temperatures 
of the He stars are typically higher than for H in this region of
parameter space.  Thus it is possible that selecting WDs on the
basis of having a low $\teffobs$ from H fits results in also 
selecting He stars with less extreme temperatures.

\subsubsection{Notes on individual objects}
\label{section:ultracoolindividual}

Three stars in the sample of candidate ultra-cool WDs
have been spectroscopically confirmed 
as being of type DC. 
SDSS~J224206$+$004822 and SDSS~J233055$+$002852 
were initially discovered by \cite{Kilic:2006} who,
from photometric fits with $\logg = 8.0$,
found $\teffobs = 3400 \unit{K}$
and $\teffobs = 4100 \unit{K}$, respectively.
Applying the Bayesian approach described in \sect{bayes} and
using the informative $\logg$ prior given in \sect{wdpop}
gives $\teffobs = 3400 \pm 30 \unit{K}$ 
for SDSS~J224206$+$004822 (where H is a better fit but neither 
model fits the data well), 
and $\teffobs = 4760 \pm 20 \unit{K}$ for 
SDSS~J233055$+$002852 (where He is clearly a better fit). 
SDSS J232115.67$+$010223.8 was reported by \cite{Carollo_etal:2006}
as being of type DC; the Bayesian results imply that this star 
most probably has a He atmosphere, with $\teffobs = 4870 \pm 20 \unit{K}$.  

\begin{figure}
\includegraphics[width=\figurewidth]
  {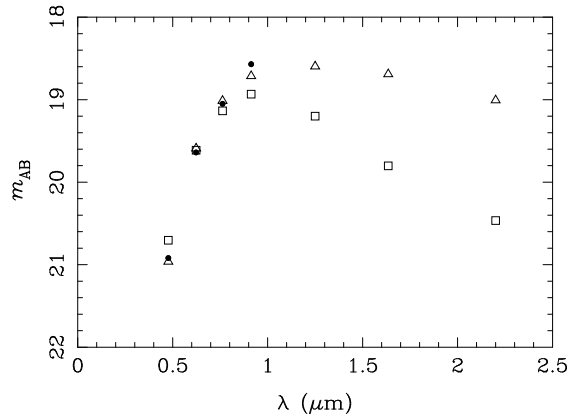}
\caption{Photometry in the 
  \gprime, \rprime, \iprime\ and \zprime bands (solid dots) 
  of the probable He WD SDSS~J205132$+$000353,
  compared to the \bestfit\ H models (open squares) and 
  He models (open triangles) in the 
  \uprime, \gprime, \rprime, \iprime, \zprime, $J$, $H$ and $K$ bands. 
  The photometric errors are $\sim 0.02$, 
  and smaller than the symbol size.}
\label{figure:ultracool8_spectrum}
\end{figure}

Only one of the \nultracool\ ultra-cool WD candidates identified by \vidrih\
was found to definitely have a He-dominated atmosphere. 
Given the strong degeneracies in the photometric WD fitting,
it was possible that this conclusion came about only due to
the sparse grid of H models `straddling' the good fit locus
while one of the He models fortuitously landed on it.
Nonetheless, these inferences are supported by the more 
thorough MCMC exploration of the model space, although the fact that neither
atmospheric species results in a good fit
(as shown in \fig{ultracool8_spectrum})
leaves the possibility
that SDSS~J205132$+$000353 is not a WD at all.
This possibility has been corroborated by Kilic (private communication),
who did not detect an appreciable proper motion from multi-epoch astrometry 
of this source, implying it is more likely to be a quasar, or other
extra-Galactic object.
These observations are at odds with proper motion of 
$\hat{\mu} = (1.02 \pm 0.01) \unit{arcsec} \unit{yr}^{-1}$ 
quoted by \vidrih;
however there are two other stars within a few arcsec of 
SDSS~J205132$+$000353, and so it is possible that the automated 
matching algorithm of \cite{Bramich_etal:2008} 
was unable to process this complicated case.

SDSS~J205132$+$000353 and
SDSS~J212930.25$-$003411.5 have very high $\vtanobs$ values
of 
$262 \pm  21 \kms$ and $224 \pm 15 \kms$, respectively.
The fact that the former is not a good fit to either
model atmosphere probably invalidates the estimate of $\vtan$,
but the latter remains a good halo candidate.

\section{Conclusions}
\label{section:conc}

WDs can be identified reliably using just photometric and astrometric 
information, but the utility of large samples generated in this 
way is limited if WDs cannot also be characterised without recourse
to spectroscopy. 
The degree to which this is possible has been assessed here by using 
Bayesian methods to include vital prior information
in the parameter estimation process.
For most WDs observed with SDSS precision photometry 
(or \percent-level photometry, but only in bands redder than $\uprime$) 
only their effective temperature 
and solid angle can be constrained unambigously.
However this very generic degeneracy can be broken by utilising
the prior information (from spectroscopic modelling) that 
most WDs have surface gravities of $7.7 \la \logg \la 8.1$.

The power of this approach was demonstrated by applying it to 
the samples of candidate halo and ultra-cool WDs generated by 
\vidrih\ from multi-epoch SDSS Stripe~82 data.
This confirms their identification of ultra-cool stars 
(as the photometric data constrain the temperature well)
but implies that most of the candidate halo stars are members of
the more numerous thick disk population 
with unusual observed colours caused by photometric noise.
This is a particular case of a very general result, that it is 
almost impossible to reliably identify a small number of outliers
if there are sufficiently many members of the more dominant population to 
scatter to the parameter space of interest.

The one situation in which photometric data are likely
to give definitive WD parameter estimates without prior information
is if \percent-level photometry is
available in the \uprime, \gprime\ and \rprime\ (or similar) bands
for H atmosphere WDs with
$10000 \unit{K} \la \teff \la 18000\unit{K}$.
As shown by \cite{Ivezic:2007}, the loci of observed colours 
of WDs in SDSS Stripe~82 exhibit an obvious separation between
H and He models (see \fig{twocolour}), and the next stage of this 
work will be to analyse a sample of such sources.
This should also have the additional benefit of giving
a more accurate surface gravity prior than used here
(in which the numbers of WDs with non-fiducial surface gravities 
are almost certainly under-estimated). 
Another important extension to this work will be to add 
a velocity prior from Galactic dynamics: for most WDs this 
will not be important, but it will be critical to obtaining
the most stringent constraints on potential halo objects.
An added side-benefit of being able to identify halo objects 
dynamically will be to place empirical 
constraints on the temperature and age distribution 
of halo WDs (\cf\ \citealt{Hansen:2001,Pauli:2006}).
Further, while the effect of the
ISM has not been included in this work
(in part to follow the assumptions of the \vidrih\ analysis), 
it would be straightforward
to include a full ISM model or Galactic extinction template 
as described in \sect{parameterest}.

These results have particularly relevance to up-coming 
large-scale deep optical surveys such as 
the SkyMapper survey \citep{Keller:2007},
the Dark Energy Survey (DES; \citealt{Flaugher_etal:2006}), 
the Panoramic Survey Telescope And Rapid Response System
(Pan--STARRS; \citealt{Kaiser:2002})
and the Large Synoptic Survey Telescope (LSST; \citealt{Ivezic:2008}).
In the context of WD studies, a critical distinction 
between these projects is that 
only SkyMapper and LSST plan to observe in the \uprime\ band.
With LSST expected to reach\footnote{Typical LSST exposures
will detect point--sources with a signal--to--noise ratio of 5
to $\uprime \simeq 24$; the depth quoted in the main text comes
from co-adding the $\sim 60$ scans LSST will make during its lifetime.
See {\tt{http://www.lsst.org/Science/lsst\_baseline.shtml}} for more details.}
$\uprime \simeq 26$ over half the sky,
it should yield a sample of $\sim 10^6$ WDs with sufficiently
precise photometric and astrometric measurements to give accurate
distances and tangential velocities.
Combined with the methods presented here, these data 
will provide an unprecedented understanding of the Galaxy's WD
population.


\section*{Acknowledgments}

Simon Vidrih provided invaluable information on the SDSS LMCC and 
the WD sample generated from it.
Many thanks to Pierre Bergeron not only for making his WD models 
publicly available, but for willingly providing extra calculations
on request. 
Mukremin Kilic provided useful insight on the first draft of this 
paper, in particular with regards to the nature of 
SDSS~J205132$+$000353.
The anonymous referee suggested a number of changes which have 
improved both the content and form of this paper.

HVP is supported by Marie Curie grant MIRG-CT-2007-203314 from the 
European Commission, and by STFC. Her work was additionally supported by 
NASA through Hubble Fellowship grant \#HF-01177.01\-A from the
Space Telescope Science Institute, which is operated by the Association of 
Universities for Research in Astronomy, Inc., for NASA, 
under contract NAS 5\-26555.  
\v{Z}I acknowledges support by NSF grant AST-0707901.


\bibliographystyle{mn2e.bst}
\bibliography{wdphoto}


\bsp

\label{lastpage}

\end{document}